\documentclass[11pt]{JHEP3}

\usepackage{latexsym}
\usepackage{epsfig,amssymb,euscript}
\usepackage{amsmath,cancel,slashed}
\DeclareMathAlphabet{\mathpzc}{OT1}{pzc}{m}{it}
\usepackage{array,calc,epsfig,dsfont}
\usepackage{multicol}
\usepackage{fancybox}
\usepackage{tensor}
\usepackage[sort]{cite}

\def\be{\begin{equation}}
\def\ee{\end{equation}}

\def\bea{\begin{eqnarray}}
\def\eea{\end{eqnarray}}
\def\bseq{\begin{subequations}}
\def\eseq{\end{subequations}}

\newcommand{\beal}{\begin{align}}

\numberwithin{equation}{section} 
\usepackage[]{graphicx}

\def\d {{\rm d}}

\def\calf         {{\cal F}}

\def\Re           {{\rm Re\hskip0.1em}}
\def\Im           {{\rm Im\hskip0.1em}}

\def\sqr#1#2{{\vcenter{\vbox{\hrule height.#2pt
 \hbox{\vrule width.#2pt height#1pt \kern#1pt \vrule width.#2pt}\hrule
 height.#2pt}}}}

  \newcommand{\sh}[1]{#1\hskip-7pt \diagup}
\newcommand{\Sh}[1]{#1\hskip-11pt \diagup}

\def\p{{\partial}}

\def\e{\epsilon}
\def\o{\omega}
\def\w{\wedge}
\def\d{\text{d}}
\def\p{\phi}
\def\ps{\psi}

\def\Re{\text{Re}}
\def\Im{\text{Im}}
\def\lrc{\lrcorner}

\def\slashchar#1{\setbox0=\hbox{$#1$}           
\dimen0=\wd0                                 
\setbox1=\hbox{/} \dimen1=\wd1               
\ifdim\dimen0>\dimen1                        
\rlap{\hbox to \dimen0{\hfil/\hfil}}      
#1                                        
\else                                        
\rlap{\hbox to \dimen1{\hfil$#1$\hfil}}   
/                                         
\fi}

\title{BPS-like potential for compactifications of heterotic M-theory?}

\author{Johannes Held${}^{\diamondsuit}$
\\

\begin{itemize}

\item  Max-Planck-Institut f\"ur Physik\\
F\"ohringer Ring 6, 80805 M\"unchen, Germany
 \end{itemize}

\bigskip
 E-mail:
\email{heldj@mppmu.mpg.de}
}

\received{\today}               
\revised{}
\accepted{}               

\preprint{hep-th\yymmddd}
\preprint{MPP-2011-108}

\abstract{We analyze the possibility to rewrite the action of Horava-Witten theory in a BPS-like form, which means that it is given as a sum of squares of the supersymmetry conditions. To this end we compactify the theory on a seven dimensional manifold of SU$(3)$ structure and rewrite the scalar curvature of the compactification manifold in terms of the SU$(3)$ structure forms. This shows that a BPS-like form cannot be obtained in general, but only for certain types of compactifications.}


\begin{document}
\newpage
\section{Introduction}
Although string theory compactifications with background fluxes are a quite old field of research, there has been renewed interest in the subject after it was recognized that one can use background fluxes for moduli stabilization \cite{Giddings:2001yu,Kachru:2003aw,Derendinger:2004jn,Villadoro:2005cu}. Even more attention was turned to flux compactification after it was understood how to use $G$-structures and generalized geometry for its description (see for example \cite{Grana:2005jc,Douglas:2006es,Blumenhagen:2006ci,Koerber:2010bx} and references therein).

In \cite{Lust:2008zd,Held:2010az} $G$-structures were used to describe vacua of type IIA and heterotic flux compactifications that break supersymmetry (SUSY). This was achieved by allowing four dimensional domain walls to be no longer BPS objects and was therefore dubbed \textit{domain wall SUSY breaking} (DWSB). It was also possible to construct explicit examples of compactification manifolds that give rise to stable DWSB vacua. An interesting question that arose is whether one can extend these examples also to the strong coupling limit.

As the strong coupling limit of type IIA and E8$\times$E8 string theory is M-theory \cite{Witten:1995ex,Horava:1995qa,Horava:1996ma,Witten:1996mz} this question can naturally be analyzed in an M-theoretic setting. Also here the use of $G$-structures turns out to be essential. The analog to Calabi-Yau (CY) compactifications in string theory would be in M-theory to compactify on manifolds with G$_2$ holonomy. But like in the CY case such an approach does not allow for non-vanishing flux. To include fluxes it is necessary to consider manifolds with G$_2$ structure instead of G$_2$ holonomy. Interestingly, seven dimensional G$_2$ structure manifolds always allow an SU$(3)$ structure \cite{Chiossi:2002tw,Bryant:2005mz} defined by an invariant one-form $v$. Identifying the direction distinguished by $v$ with the extra dimension that becomes compact by going from M- to  string theory suggests that one can compare M-theory compactifications on seven dimensional SU$(3)$ structure manifolds to string compactifications with six dimensional SU$(3)$ structure \cite{Candelas:1984yd,Kaste:2003zd,Dall'Agata:2003ir,Behrndt:2003zg,Behrndt:2003uq,Behrndt:2004bh,House:2004pm,Behrndt:2005im,Micu:2006ey,Anguelova:2006qf}.

To arrive at the DWSB solutions in \cite{Lust:2008zd,Held:2010az} the action was written in terms of an effective scalar potential. It was then shown with the help of the Bianchi identity for the flux that this potential can be brought into a BPS-like form, which means that the potential consists only of terms that are squares of SUSY conditions. Differently stated a BPS-like form makes it explicit that the Bianchi identity and SUSY guarantee that the equations of motion (EoM's) are satisfied. So, a first step in the analysis of the strong coupling extension of the results of \cite{Lust:2008zd,Held:2010az} would be to establish a BPS-like potential for M-theory.

We will study in this paper whether such a BPS-like potential is available for the low-energy limit of heterotic M-theory, i.e. for eleven dimensional supergravity living on a space with two boundaries. Indeed, we will show that \textit{in general it is not possible to bring the action of Horava-Witten theory into a BPS-like form}. However, we will show that a BPS-like form is available for a large class of seven dimensional compactification manifolds and also that a reduction to ten dimensions gives back the results of \cite{Held:2010az}.

This paper is organized as follows. In section~\ref{sec:Mreview} we give a short review of heterotic M-theory. In section~\ref{sec:ScalPotAndEOM} we derive an effective scalar potential and show that the EoM's derived from this potential are equivalent to those obtained from the original action. Section~\ref{sec:G2AndSU3Structure} gives the most important facts on G$_2$ and SU$(3)$ structures in seven dimensions. In section~\ref{sec:RG2} we rewrite the scalar curvature $R$ of the compactification manifold in terms of the SU$(3)$ structure forms. The SUSY conditions for our compactification of Horava-Witten theory are reviewed and newly analyzed in section~\ref{sec:Supersymmetry conditions}. In section~\ref{sec:BPSPot?} the results are gathered in order to show that a BPS-like potential is not possible in general. The conditions under which such a potential can be achieved are also given there. We close the paper by discussing various limits in order to give crosschecks for our previous results. These crosschecks all turn out to be successful. Most importantly we show that our ansatz gives back the results of \cite{Held:2010az} once reduced to ten dimension. Our conventions are summarized in appendix~\ref{app:conventions}. Appendix~\ref{app:Curvature} gives more details on the calculation of $R$ in terms of G$_2$ structure invariants and appendix~\ref{app:SUSYconstraints} gives the complete list of SUSY conditions on the flux we found.
\section{A short review of heterotic M-theory}
\label{sec:Mreview}
As was shown by Horava and Witten \cite{Horava:1995qa,Horava:1996ma} the strong coupling limit of heterotic string theory can be described as eleven dimensional supergravity with two ten dimensional boundaries. The action of this theory can be split into a bulk and a boundary part. The bulk action is the standard action of eleven dimensional supergravity \cite{Cremmer:1978km}
\begin{equation}
\label{eq:11daction}
S_0~=~\frac{1}{2\,\kappa^2}\,\int\limits_{X_{11}}{\text{dvol}_{11}\,(R^{(11)}\,-\,2|G_{11}|^2)}\,-\,\frac{2}{3\,\kappa^2}\,\int\limits_{X_{11}}{C_{11}\w G_{11}\w G_{11}}~,
\end{equation}
where $C_{11}$ is the three-form potential for the four-form flux $G_{11}$: $\d C_{11}=G_{11}$. In addition to the fields in the bulk there is a gauge field contribution on each boundary given by
\begin{equation}
\label{eq:11dboundaryaction}
S_b~=~-\,\frac{1}{8\pi\kappa^2}\Big(\frac{\kappa}{4\pi}\Big)^{2/3}\sum\limits_{p=1,2}\,\int\limits_{B_{10,p}}{\text{dvol}_{10,p}\,\Big(\text{Tr}\calf_{p}^2\,-\,\frac{1}{2}\text{Tr}(R^{(10)}_{p})^2\Big)}~.
\end{equation}
The two-forms $\calf_{p}$ are $E_8$ field strengths and the trace Tr over the gauge fields is related to the trace in the adjoint representation by Tr $=\frac{1}{30}\text{tr}_{\text{adj}}$. The trace of the curvature two-forms is defined to be\footnote{For a list of our notational conventions, see appendix \ref{app:conventions}.}
\begin{equation}
\text{Tr}(R^{(10)}_{p})^2~:=~-\,(R_p^{(10)})\indices{^I_J}\lrc(R_p^{(10)})\indices{^J_I}~=~\frac{1}{2}(R_p^{(10)})_{IJKL}(R_p^{(10)})^{IJKL}~.
\end{equation}
One should note that the boundary terms come with an extra factor of $\kappa^{2/3}$ and lead to a perturbation of the bulk theory. In particular $\d C_{11}$ is only the leading contribution to $G_{11}$ and it is not possible to set $G_{11}$ to zero identically. Also the Bianchi identity for $G_{11}$ gets contributions from the two boundaries.

For our later calculations it is worthwhile to note that one can view Horava-Witten theory from two different perspectives. In the so called upstairs picture one considers as eleventh dimension the orbifold $S_1/\mathbb{Z}_2$ and identifies $x^{11}\sim x^{11}+2\pi\rho\sim-x^{11}$. This gives two fixed ten dimensional hyperplanes at $x_1^{11}=0$ and $x_2^{11}=\pi\rho$ on which the action $S_b$ lives. The Bianchi identity reads then up to $\mathcal{O}(\kappa^{2/3})$
\begin{equation}
\label{eq:GBI}
\frac{1}{4!}(\d G_{11})_{IJKL\,11}\d x^{IJKL}~=~-\frac{1}{8\pi}\Big(\frac{\kappa}{4\pi}\Big)^{2/3}\sum\limits_{i=1,2}\delta(x^{11}-x_i^{11})(\text{Tr}\calf_i\w\calf_i\,-\,\frac{1}{2}\text{Tr}R^{(10)}_i\w R^{(10)}_i)~,
\end{equation}
and the four-form flux $G$ is fixed to be
\begin{align}
\label{eq:FourFlux}
&(G_{11})_{IJKL}&&=&&-\,\frac{1}{16\pi}\Big(\frac{\kappa}{4\pi}\Big)^{2/3}\,\Big[\epsilon(x^{11})\, K^{(1)}\,-\,\frac{x^{11}}{\pi\rho}(K^{(1)}\,+\,K^{(2)})\Big]_{IJKL}&\\
&(G_{11})_{IJK\,11}&&=&&2\,\partial_{[I}(C_{11})_{JK]\,11}\,-\,\frac{1}{16\pi^2\rho}\Big(\frac{\kappa}{4\pi}\Big)^{2/3}\Big[\omega^{(1)}_3\,+\,\omega^{(2)}_3\Big]_{IJK}&\nonumber
\end{align}
for $x\in\big[-\pi\rho,\,\pi\rho\big]$. The step function $\epsilon(x)$ is $-1$ for $x<0$ and $1$ for $x>0$ and the $K^{(p)}$ are defined to be
\begin{equation}
K^{(p)}~=~\d\omega^{(p)}_3~=~\text{Tr}\calf_p\w\calf_p\,-\,\frac{1}{2}\text{Tr}R^{(10)}_p\w R^{(10)}_p~.
\end{equation}

In the downstairs picture one takes as eleventh dimension an interval of length $\pi\rho$ with the hyperplanes as boundaries. The action $S_0$ has then to be supplemented with appropriate boundary conditions for the fields \cite{Horava:1996ma,Lukas:1998ew}.

In order to obtain heterotic supergravity one has to reduce the bulk action dimensionally and take the limit $\rho\rightarrow 0$. Then the flux $(G_{11})_{IJKL}$ becomes zero and the two pieces of the M-theory boundary action will combine to give the $\mathcal{O}(\alpha')$ terms of heterotic supergravity. 

In order to make contact to the results presented in \cite{Held:2010az}, one has to deal with two subtleties. Firstly, in heterotic flux compactifications it is useful to use a torsionful connection $\omega_+$ in order to construct the Tr$R^2$ term (see e.g. \cite{Bergshoeff:1989de,Lopes Cardoso:2003af,Becker:2009zx}), while in M-theory one uses the Levi-Civita connection. However, as the connection is not essential for anomaly cancellation \cite{Hull:1985dx,Hull:1986kz}, it is possible to use the torsionful connection also in heterotic M-theory. Thus one should replace $R$ with $R_+=R(\omega_+)$ in the boundary action (\ref{eq:11dboundaryaction}) and the Bianchi identity (\ref{eq:GBI}) if one wants to make contact to heterotic flux compactifications. Secondly, in order to put the action into string frame one has to perform a Weyl transformation of the metric (see e.g. \cite{Becker:2007zj,Polchinski:1998rr}). This will lead to additional contributions coming from the Tr$R_+^2$ term. Since these terms are of fourth order in derivatives and are not necessary to ensure anomaly cancellation, we will consider them as higher order contributions and neglect them in our discussion.

\section{Scalar potential and equations of motion}
\label{sec:ScalPotAndEOM}
We are interested in the question whether it is possible to rewrite the action of Horava-Witten theory in a BPS-like form. Since in the end we want to consider compactifications to four dimensions, we will rewrite the action as a four dimensional integral over an effective potential along the lines of \cite{Held:2010az,Lust:2008zd}. We then show that the equations of motion derived from both formulations are equivalent. Therefore it will be sufficient to check whether the scalar potential can be brought to a BPS-like form.
\subsection{The scalar potential}
\label{sec:TheScaPot}
Considering compactifications respecting four dimensional Poincare invariance metric and flux should be decomposed as
\begin{equation}
\label{eq:metricMto4}
\d s_{11}^2~=~e^{2A}\d \hat{s}^2_4\,+\,\d s_7^2~,
\end{equation}
\begin{equation}
C_{11}~=~\frac{1}{3!}\,(C^{(4)})_{\mu\nu\sigma}\d x^{\mu\nu\sigma}\,+\,\frac{1}{3!}\,C_{mnp}\d x^{mnp}~,
\end{equation}
\begin{equation}
\label{eq:G11decomposition}
G_{11}~=~\tilde{\mu}\,\hat{\text{dvol}}_4\,+\frac{1}{4!}\,G_{mnpq}\d x^{mnpq}~,
\end{equation}
with $A$ being the warp factor, $\d\hat{s}^2_4$ the metric of a maximally symmetric four dimensional space-time, and $\d s^2_7$ the metric of the seven dimensional manifold $M$. $\tilde{\mu}$ is a real constant and $G$ is the four form flux in the seven internal dimensions. Then the bulk action becomes
\begin{align}
&S_0&&=~\frac{1}{2\,\kappa^2}\,\int\limits_{X_4}\hat{\text{dvol}}_4\int\limits_{M}\text{dvol}_7\Big[e^{2A}\,\hat{R}^{(4)}\,+\,e^{4A}\,(R\,-\,8\,\nabla^2 A\,-\,20\,\d A^2\,-\,2|G|^2\nonumber\\
&&&~-\,2\,\mu^2\,-\,4\mu\,C\lrcorner \ast G)\Big]&\\
&&&=~\frac{1}{2\,\kappa^2}\,\int\limits_{X_4}\hat{\text{dvol}}_4{\int\limits_{M}\text{dvol}_7{\left[e^{2A}\,\hat{R}^{(4)}\,-\,\mathcal{V}\right]}}~=~-\,\int\limits_{X_4}{\hat{\text{dvol}}_4 \,V_0}\,.&\nonumber
\end{align}
Here $R$ denotes the Ricci scalar of $M$ and $\mu=e^{-4A}\tilde{\mu}$. All fields except the unwarped four dimensional curvature $\hat{R}^{(4)}$ depend only on the seven dimensional internal space $M$. 

On the boundary the metric splits into a ten and a one dimensional piece
\begin{equation}
\label{eq:MetricNearBound}
\d s_{11}^2~=~\d s_{10}^2\,+\,v\otimes v~,
\end{equation}
with $\d s_{10}^2$ the metric on the boundary. As we explained in section \ref{sec:Mreview} a rescaling of $g_{10}\rightarrow g'_{10}= e^{-\sigma} g_{10}$, which is necessary to reach the string frame, will only introduce terms with four derivatives, which we neglect in our analysis. The metric $g'_{10}$ is compactified according to
\begin{equation}
\d (s'_{10})^{2}~=~e^{2A'}\d \hat{s}_4^2\,+\,\d (s'_{6})^{2}~,
\end{equation}
where $A'$ is a shifted warp factor $A'=A+\frac{1}{2}\sigma$. Taking this into account, the boundary action can be written as
\begin{align}
&S_b~=~-\,\frac{1}{8\pi\kappa^2}\Big(\frac{\kappa}{4\pi}\Big)^{2/3}\sum\limits_{p=1,2}\,\int\limits_{X_4}\hat{\text{dvol}}_4\int\limits_{B_{6,p}}{\text{dvol}'_{6,p}\,e^{4A'-3\sigma}\,\Big\lbrace\Big(\text{Tr}(\calf^{(6)}_{p})^2\,-\,\frac{1}{2}\text{Tr}(R^{(6)'}_{p\,+})^2\Big)}\nonumber\\
&~~~~-\,\frac{1}{24}\big|e^{-2A'}\,\hat{R}^{(4)}-12|\d A'|^2\big|^2-4\,e^{-2A'}\,(\nabla_i\nabla_j e^{A'})(\nabla^i\nabla^j e^{A'})-8\,\big|\d A'\lrc H\big|^2\Big\rbrace~.
\end{align}
Here we used that the $E_8$ gauge fields are confined to the internal space $\calf=\calf^{(6)}$. The appearing three form field $H_{ijk}=G_{11\,ijk}$ was used to construct the torsionful curvature tensor $R_+$ in the same way as the Neveu-Schwarz three form is used as torsion in the heterotic string. Note that we kept four derivative terms in this expression although we discarded them in our previous discussion. We do this only in order to compare to the results found in \cite{Held:2010az} for the heterotic string.

The action can thus be written as a four dimensional integral over a scalar potential
\begin{equation}
\label{eq:potential0}
S~=~-\int\limits_{X_4}\hat{\text{dvol}}_4 \,V~=~-\int\limits_{X_4}\hat{\text{dvol}}_4 \,(V_0\,+\,V_b)
\end{equation}
combining the contribution from the bulk $V_0$ and from the boundary $V_b$, respectively, which are given by
\begin{subequations}
\begin{align}
\label{eq:pot0bulk}
&V_0~=~\frac{1}{2\,\kappa^2}\,\int\limits_{M}\text{dvol}_7{\Big[-\,e^{2A}\,\hat{R}^{(4)}\,-\,e^{4A}\,(R-8\,\nabla^2 A-20\,\d A^2-2|G|^2-2\mu^2-4\mu\,C\lrcorner \ast G)\Big]}\\
\label{eq:pot0boundary}
&V_b~=~\frac{1}{8\pi\kappa^2}\Big(\frac{\kappa}{4\pi}\Big)^{2/3}\sum\limits_{p=1,2}\,\int\limits_{B_{6,p}}{\text{dvol}'_{6,p}\,e^{4A'-3\sigma}\,\Big\lbrace\Big(\text{Tr}(\calf^{(6)}_{p})^2\,-\,\frac{1}{2}\text{Tr}(R^{(6)'}_{p\,+})^2\Big)}\\
&~~~~-\,\frac{1}{24}\big|e^{-2A'}\,\hat{R}^{(4)}-12|\d A'|^2\big|^2-4\,e^{-2A'}\,(\nabla_i\nabla_j e^{A'})(\nabla^i\nabla^j e^{A'})-8\,\big|\d A'\lrc H\big|^2\Big\rbrace~.\nonumber
\end{align}
\end{subequations}
\vskip11pt

%
%
\subsection{Equations of motion}
We will now show that the equations of motion derived from (\ref{eq:potential0}) are consistent with the full eleven dimensional equations coming from (\ref{eq:11daction}) and (\ref{eq:11dboundaryaction}). In order to proof this we will make use of the downstairs picture. Setting the variation of the fields at the boundaries to zero, as is usual, we are left with the bulk action plus boundary conditions \cite{Lukas:1998ew}. Since these boundary conditions will not change in going from eleven to four dimensions we only have to consider the bulk part of (\ref{eq:potential0}).

Varying (\ref{eq:11daction}) with respect to the metric $g^{(11)}$ and the three form potential $C_{11}$ one obtains
\begin{align}
&\delta_{g^{(11)}}:&&R^{(11)}_{MN}\,-\,\frac{1}{2}\,g^{(11)}_{MN}\Big[R^{(11)}\,-\,2\,|G_{11}|^2\Big]-2\,(\iota_M G_{11})\lrcorner(\iota_N G_{11})~=~0&\label{eq:11deomg}\\
&\delta_{C_{11}}:&&\d\ast_{11} G_{11}\,+\,G_{11}\w G_{11}~=~0~.&\label{eq:11deomC}
\end{align}
Restricting (\ref{eq:11deomg}) to internal coordinates $MN=mn$ and inserting (\ref{eq:G11decomposition}) leads to
\begin{align}
&R_{mn}\,-\,4\,e^{-A}\nabla_m\nabla_n e^{A}\,-\,2\,\iota_m G\lrcorner \iota_n G&\\
&\,-\,\frac{1}{2}\,g_{mn}\Big[e^{-2A}\,\hat{R}^{(4)}\,+\,R\,-\,8\,\nabla^2\,A\,-\,20\,\d A^2\,-\,2\mu\,-\,2\,|G|^2\Big]~=~0~.&\nonumber
\end{align}
Taking the trace of (\ref{eq:11deomg}) over its external indices one finds
\begin{align}
&6\,\nabla^2e^{2A}~=~\hat{R}^{(4)}\,+\,2\,e^{2A}\big(R\,+\,2\mu\,-\,2|G|^2\big)~.&
\end{align}
Furthermore, inserting (\ref{eq:G11decomposition}) into the equation of motion for $C_{11}$ (\ref{eq:11deomC}) gives
\begin{equation}
\d\ast G~=~-\,2\mu\,G~.
\end{equation}
These are exactly the EoM's that one obtains if one varies the bulk part of  (\ref{eq:potential0}) with respect to the internal metric $g^{mn}$, the warp factor $A$, and the flux $G$, respectively. Thus by inserting our compactification ansatz into the equations of motion of the eleven dimensional theory we obtain the same results as if we vary (\ref{eq:potential0}). We will therefore work with the effective potential instead of the full eleven dimensional action.
\section{G$_2$ and SU$(3)$ structure for seven dimensional manifolds}
\label{sec:G2AndSU3Structure}
We will review in this section some points concerning G$_2$ and SU$(3)$ structure on seven dimensional manifolds which will become important later on (see e.g. \cite{Chiossi:2002tw,Bryant:2005mz,Kaste:2003zd,Dall'Agata:2003ir,Behrndt:2003zg,Behrndt:2003uq,Behrndt:2004bh,House:2004pm,Behrndt:2005im,Micu:2006ey,Anguelova:2006qf}). A G$_2$ structure manifold is completely determined by its invariant three form $\p$, or equivalently by a globally well defined SO$(7)$ Majorana spinor $\eta$. Normalizing this spinor such that $\| \eta \|^2=1$ these quantities are related by
\begin{align}
\label{eq:PhiAndPsi}
&\p~=~\frac{i}{3!}\,\eta^{\dagger}\,\gamma_{mnp}\,\eta\,\d x^{mnp}&&\ast\p~=~\ps~=~-\frac{1}{4!}\,\eta^{\dagger}\,\gamma_{mnpq}\,\eta\,\d x^{mnpq}~.&
\end{align}
For a manifold of G$_2$ holonomy $\d\p=\d\ps=0$ would hold. The departure from holonomy can be measured by the G$_2$ torsion classes
\begin{align}
\label{eq:dphiG2torsion}
&\d\p~=~\tau_0\,\ps+3\,\tau_1\w\p+\ast\,\tau_3\,,&&\d\ps~=~4\,\tau_1\w\ps+\tau_2\w\p~.&
\end{align}
Inverting (\ref{eq:dphiG2torsion}) it is possible to express the torsion classes in terms of $\p$ and $\ps$
\begin{align}
\label{eq:G2torsion}
&\tau_0~=~\frac{1}{7}\,\d\p\lrcorner\ps\,,&&\tau_1~=~-\frac{1}{12}\,\d\p\lrcorner\p~=~\frac{1}{12}\,\d\ps\lrcorner\ps\,,&\\
&\tau_2~=~\frac{1}{2}\left(\d\ps\lrcorner\p-\ast\,\d\ps\right)-2\,\tau_1\lrcorner\p&&\tau_3~=~\ast\d\p-\tau_0\,\p+3\,\tau_1\lrcorner\ps\,.&\nonumber\\
&~~~\;=~-\ast\,\d\ps+4\,\tau_1\lrcorner\p\,,&&&\nonumber
\end{align}
An SU$(3)$ structure can be obtained with the help of a globally defined invariant one form $v$, either by suitable contractions of $\p$ and $\ps$ with $v$, or due to a modification of the spinor $\eta$. Since this spinor will appear also later in the SUSY transformations we choose the second description and define
\begin{align}
\label{eq:etaplus}
&\eta_+~=~\frac{1}{\sqrt{2}}\;e^{\frac{Z}{2}}(1+v_m\gamma^m)\,\eta\,,&&\eta_+^*~=~\eta_-~=~\frac{1}{\sqrt{2}}\;e^{\frac{Z}{2}}(1-v_m\gamma^m)\,\eta~.&
\end{align}
Here $Z$ is a real function and $v_m$ denotes the components of $v$. With these two spinors one can construct several new forms on $M$
\begin{align}
&\Sigma_p~=~\frac{1}{p!}\;\eta_+^{\dagger}\,\gamma_{n_1\ldots n_p}\,\eta_+\,\d x^{n_1\ldots n_p}\,,&&\tilde{\Sigma}_p~=~\frac{1}{p!}\;\eta_+^{T}\,\gamma_{n_1\ldots n_p}\,\eta_+\,\d x^{n_1\ldots n_p}~.&
\end{align}
From these the forms $\Sigma_2$ and $\tilde{\Sigma}_3$ can be related to the SU$(3)$ structure forms $J$ and $\Omega$, while $\Sigma_1$ is proportional to $v$
\begin{align}
\label{eq:SigmaSU3}
&v~=~e^{-Z}\,\Sigma_1\,,&&J~=~i\,e^{-Z}\Sigma_2\,,&&\Omega~=~i\,e^{-Z}\tilde{\Sigma}_3~.
\end{align}
A detailed calculation shows that $J$ and $\Omega$ satisfy indeed the SU$(3)$ relations
\begin{align}
&J\w \Omega~=~0\,,&&\text{dvol}_7~=~v\w\text{dvol}_6~=~\frac{1}{3!}\,v\w J\w J\w J~=~-\frac{i}{8}\,v\w \Omega\w\bar{\Omega}~,
\end{align}
and that $\Omega$ is a $(3,0)$ form with respect to the almost complex structure defined by $J$
\begin{equation}
J\indices{_m^n}\,\Omega_{npq}~=~-i\,\Omega_{mpq}~.
\end{equation}
Furthermore, $v$ is perpendicular to $J$ and $\Omega$ and thus $M$ looks locally like the direct product of an SU$(3)$ structure manifold and a line. Writing the G$_2$ spinor $\eta$ in terms of $\eta_+$ one gets from (\ref{eq:PhiAndPsi}) the connection between the G$_2$ and the SU$(3)$ forms
\begin{align}
\label{eq:PhiPsiDecompositons}
&\p~=~v\w J+\text{Re}\Omega\,,&&\ps=\frac{1}{2}\,J\w J+v\w \text{Im}\Omega~.&
\end{align}
Like in the G$_2$ case the departure from SU$(3)$ holonomy is measured by torsion classes.
\begin{align}
\label{eq:SU3TorsionClasses}
&\d v&&=&&~~~R\,J\,+\,\bar{V}_1\lrcorner\,\Omega\,+\,V_1\lrcorner\,\bar{\Omega}\,+\,v\w W_0\,+\,T_1&\\
&\d J&&=&&-\frac{3}{2}\,\text{Im}(\bar{W}_1\,\Omega)\,+\,W_4\w J\,+\,W_3\,+\,v\w\Big(\,\frac{2}{3}\,\text{Re}E\,J\,+\,\bar{V}_2\lrcorner\,\Omega\,+\,V_2\lrcorner\,\bar{\Omega}\,+\,T_2\Big)&\nonumber\\
&\d\Omega&&=&&~~~W_1\,J\w J\,+\,W_2\w J\,+\,\bar{W}_5\w\Omega\,+\,v\w\,(E\,\Omega\,-\,4V_2\w J\,+\,S)~.&\nonumber
\end{align}
Here $R$ is a real scalar, while $W_1$ and $E$ are complex scalars. $W_5$, $V_1$, and $V_2$ are $(1,0)$ forms, while $W_0$ and $W_4$ are real one forms. $W_2$, $T_1$, $T_2$ are primitive and $(1,1)$. $W_3$ and $S$ are $(2,1)+(1,2)$ and primitive. All degrees of the forms are understood with respect to the almost complex structure defined by $J$ and $\Omega$. Note that while $W_1$ to $W_5$ are also present in the six dimensional case, the other torsion classes are special to seven dimensions and describe the embedding of the SU$(3)$ structure manifold into $M$. With this tools at hand we turn now back to the effective action.
\section{The Ricci scalar of G$_2$ manifolds}
\label{sec:RG2}
A  main obstacle in the analysis of (\ref{eq:potential0}) is the seven dimensional Ricci scalar $R$. Fortunately all the information encoded in the metric $g$ of a $G$-structure manifold is also contained in the forms invariant under $G$. Therefore one can express the Ricci scalar equivalently well in terms of these forms, as was done in \cite{Lust:2008zd,Held:2010az} in the context of string compactifications to four dimensions. In this section we are going to show how this works for compactifications on seven dimensional manifolds with G$_2$ structure.
\subsection{$R$ in terms of G$_2$ structure}
The Ricci scalar for G$_2$ structure manifolds was worked out by Bryant \cite{Bryant:2005mz} in terms of torsion classes
\begin{equation}
R~=~-12\,\ast\d\ast\tau_1+\frac{21}{8}\,\tau_0^2+30\,|\tau_1|^2-\frac{1}{2}\,|\tau_2|^2-\frac{1}{2}\,|\tau_3|^2~.
\end{equation}
Due to the complicated dependence of $\tau_2$ and $\tau_3$ on $\p$ and $\ps$ given in (\ref{eq:G2torsion}) this seems not to be a very pleasing formula. But with a little algebra that is given in appendix \ref{app:Curvature} it is possible to show that the absolute values of the torsion classes are not independent
\begin{align}
&|\tau_2|^2~=~|\d\ps|^2\,-\,48\,|\tau_1|^2\\
&|\tau_3|^2~=~|\d\p|^2\,-\,36\,|\tau_1|^2\,-\,7\,|\tau_0|^2~.\nonumber
\end{align} 
Thus the scalar curvature $R$ can be written as
\begin{align}
\label{eq:scalcurv}
R~&=~12\,\delta\tau_1+\frac{49}{8}\tau_0^2+72\,|\tau_1|^2-\frac{1}{2}\,|\d\p|^2-\frac{1}{2}\,|\d\ps|^2\\
&=~-\nabla^m\left(\d\ps\lrcorner\ps\right)_m+\frac{1}{2}\,|\d\ps\lrcorner\ps|^2+\frac{1}{8}\,|\d\p\lrcorner\ps|^2-\frac{1}{2}\,|\d\p|^2-\frac{1}{2}\,|\d\ps|^2\nonumber
\end{align}
and depends only on $\p$, $\ps$ and their exterior derivatives.
\subsection{$R$ in terms of SU$(3)$ structure}
Although equation (\ref{eq:scalcurv}) provides a good description for $R$ as a function of $\p$ and $\ps$ this form is not convenient for our purposes as we are interested in manifolds with SU$(3)$ structure. The next task is thus to decompose $\p$ and $\ps$ according to (\ref{eq:PhiAndPsi}) and find the expression for $R$ in terms of $v$, $J$, and $\Omega$. A lengthy calculation provides the building blocks of $R$
\begin{align}
\label{eq:Rbuilding}
&\big|\d\p\big|^2~=~\big|v\w\d J-\Re\,\d\Omega\big|^2\,-\,\big|\d v\w J-\Re\,\d\Omega\big|^2+\big|\Re\,\d\Omega\big|^2+2\,\big|\d v\w J\big|^2&\\
&~~~~~~~~~+\big|\d v\lrc v-\d J\lrc J\big|^2-\big|\d v\lrc v\big|^2-\big|\d J\lrc J\big|^2\,,&\nonumber\\
&&\nonumber\\
&\big|\d\ps\big|^2~=~\Big|\frac{1}{2}\,\d J^2+\d v\w\Im\,\Omega\Big|^2-\Big|\Im\,\d\Omega\lrc v\Big|^2-\Big|\d v\lrc v\Big|^2-\Big|\frac{1}{2}\,\d J^2\lrc v\Big|^2&\\
&~~~~~~~~~~~+\Big|\d v\lrc v+\Im\,\d\Omega\lrc\Im\,\Omega\Big|^2-\Big|\Im\,\d\Omega\lrc\Im\,\Omega\Big|^2+\Big|\frac{1}{2}\,\d J^2\lrc v-\Im\,\d\Omega\Big|^2\,,&\nonumber\\
&&\nonumber\\
&\Big|\d\ps\lrc\ps\Big|^2~=~\Big|\frac{1}{4}\d J^2\lrc J^2-\Re\,\d\Omega\lrc(v\w J)+\d v\lrc\Im\,\Omega-2\d v\lrc v-\Im\,\d\Omega\lrc\Im\,\Omega\Big|^2&\\
&~~~~~~~~~~~~~~+\frac{1}{4}\,\Big|\d J^2\lrc(v\w J^2)\Big|^2-\Big|\Im\,\d\Omega\lrc(v\w\Im\,\Omega)\Big|^2+\frac{1}{16}\Big|\Im\,\d\Omega\lrc(J^2+2v\w\Im\,\Omega)\Big|^2&\nonumber\\
&~~~~~~~~~~~~~~+\frac{1}{4}\,\Big|\Im\,\d\Omega\lrc J^2\Big|^2-\frac{1}{16}\,\Big|2\d J^2\lrc(v\w J^2)+\Im\,\d\Omega\lrc(J^2+2v\w\Im\,\Omega)\Big|^2\,,&\nonumber\\\
&&\nonumber\\
&\Big|\d\p\lrcorner\,\ps\Big|^2~=~\Big|2\d v\lrcorner J+\frac{1}{2}\,\text{Re}\,\d\Omega\lrcorner\,J^2-\d J\lrcorner \text{Im}\,\Omega+\text{Re}\,\d\Omega\lrcorner (v\w \text{Im}\,\Omega)\Big|^2&\\
&~~~~~~~~~~\,\,~=~\Big|2\d v\lrcorner J+\text{Re}\,\d\Omega\lrcorner\,J^2-\frac{1}{2}\,\Im[ \d\Omega\lrc(v\w\bar{\Omega})]\Big|^2\,,&\nonumber\\
&&\nonumber\\
&\d\ps\lrcorner\,\ps~=~\frac{1}{4}\d J^2\lrcorner\,J^2+\d v\lrc\Im\Omega-\frac{1}{2}v(\text{Im}\,\d\Omega\lrcorner J^2)-\Re\,\d\Omega\lrc(v\w J)&\\
&~~~~~~~~~~~~~~~-2\d v\lrc v - \Im\,\d\Omega\lrc\Im\Omega-v\,\big(\Im\,\d\Omega\lrc(v\w\Im\Omega)\big)\,,&\nonumber
\end{align}
which are at this stage not very illuminating.\footnote{Note that $\d J^2\lrc(v\w \Im\Omega)=-2\,\Re\,\d\Omega\lrc(v\w J)$.} Most of the appearing parts are square terms, but there are also a lot of linear contributions coming from $\d\ps\lrc\ps$. Also, in order to check for a BPS-like potential we have to know whether the square terms vanish if we impose supersymmetry. To this end we turn in the next section to the investigation of the SUSY conditions.
\section{Supersymmetry conditions}
\label{sec:Supersymmetry conditions}
Eleven dimensional supergravity comes with one Majorana spinor $\e$ as SUSY generator. In Horava-Witten theory the SUSY variations of the gravitino $\psi_M$ and the gauginos $\chi_p$ on each boundary are given by
\begin{subequations}
\begin{equation}
\delta\Psi_M\,=\,\left[\nabla_M+\frac{1}{144}\,\left(\Gamma\indices{_M^{NPQR}}-8\,\delta_M^N\,\Gamma^{PQR}\right)(G_{11})_{NPQR}\right]\e\,,
\end{equation}
\begin{equation}
\delta\chi_p\,=\,-\frac{1}{4}\big(\Gamma^{IJ}\calf_{p\,IJ}\big)\e~.
\end{equation}
\end{subequations}
In \cite{Behrndt:2005im} it was shown that there are three ways to decompose the spinor $\e$ such that one obtains an SU$(3)$ invariant SO$(7)$ spinor that can be identified with $\eta_+$. The possibilities are further restricted to two if one wants to consider $\mathcal{N}=1$ SUSY in four dimensions. Finally, to describe heterotic M-theory one is restricted to use a decomposition into a chiral 4d spinor $\chi_+$ and $\eta_+$
\begin{equation}
\e~=~\chi_+\otimes\eta_++\chi_-\otimes\eta_-~=~\chi_+\otimes\eta_++\,c.\,c.~.
\end{equation}
Using this split the gaugino variations can be rewritten with the help of (\ref{eq:SigmaSU3}) and yield the well known conditions that $\calf_p$ is $(1,1)$ and primitive with respect to $J$
\begin{align}
\label{eq:Fcond}
&\calf_p\lrc J~=~0& &\calf_p^{2,0}~=~\calf_p^{0,2}~=~0~.&
\end{align}
The eleven dimensional gravitino variation gives rise to two sets of conditions\footnote{Here we used the AdS killing spinor equation $\hat{\nabla}_{\mu}\chi_+=1/2\,w_0^*\hat{\gamma}_{\mu}\chi_-$.}
\begin{subequations}
\label{eq:SUSYcond}
\begin{align}
&\delta\Psi_{\mu}=0~\Rightarrow~e^{-A}w_0\,\eta_+^*+\left(\sh{\partial}A+\frac{1}{3}\sh{G}+\frac{2i\mu}{3}\right)\eta_+~=~0\label{eq:exSUSYcond}\\
&\delta\Psi_{m}=0~\Rightarrow~\nabla_m\,\eta_+~=~\frac{1}{144}\,\left(i\mu\gamma_m+8\,G_{mpqr}\gamma^{pqr}-G_{npqr}\gamma\indices{_m^{npqr}}\right)\eta_+~.\label{eq:inSUSYcond}
\end{align}
\end{subequations}
The first of these equations will give algebraic constraints on the flux $G_{11}$. Easily to see is that a contraction of (\ref{eq:exSUSYcond}) with $\eta_+^{\dagger}$ leads to $\mu=0$. Therefore we will consider only internal four flux and set $\mu=0$ in what follows. The second one translate into differential conditions on $v$, $J$, and $\Omega$. Similar analyses have also been performed by \cite{Kaste:2003zd,Dall'Agata:2003ir,Behrndt:2003zg,Behrndt:2003uq,Behrndt:2004bh,Behrndt:2005im}, whose results are equivalent to our results.
\subsection{Differential conditions}
Contracting (\ref{eq:inSUSYcond}) with $\eta_+^{\dagger}\gamma_{n_1\ldots n_{p-1}}$ and anti-symmetrizing over all indices gives the exterior derivatives of $\Sigma_{p}$. Exchanging $\eta_+^{\dagger}$ with $\eta_+^T$ yields the derivatives of $\tilde{\Sigma}_{p}$, which can be converted with (\ref{eq:SigmaSU3}) into the derivatives of $v$, $J$, $\Omega$, and their wedge products. Furthermore, $\d Z= \d A$ and $\d(v\w J)=\d v\w J-v\w\d J$ demands that $w_0=0$. For supersymmetric vacua we are thus dealing with compactifications to warped Minkowski space that obey only internal flux and whose internal manifold has to satisfy the conditions\footnote{To obtain these simple expressions one also have to make use of (\ref{eq:exSUSYcond}).}
\begin{subequations}
\label{eq:DiffSUSYcond}
\begin{align}
&e^{-2A}\d\big(e^{2A}\,v\big)&&=&&\phantom{-}0&\label{eq:DiffSUSYv}\\
&e^{-3A}\d\big(e^{3A}\Omega\big)&&=&&\phantom{-}0&\\
&e^{-4A}\d\big(e^{4A}\,J\big)&&=&&\phantom{-}2\,\ast G&\\
&e^{-2A}\d\big(e^{2A}\,J\w J\big)&&=&&-4\,v\w G\\
&\d\big(J\w J\w J\big)&&=&&-12\,v\w J\w G~.
\end{align}
\end{subequations}
\subsection{Conditions on the flux}
Acting on (\ref{eq:exSUSYcond}) with $\eta_+^{\dagger}\gamma_{n_1\ldots n_{p-1}}$ and  $\eta_+^{T}\gamma_{n_1\ldots n_{p-1}}$ gives various constraints on the flux which we listed in appendix \ref{app:SUSYconstraints}. Most important of these are the three restrictions
\begin{align}
\label{eq:Gconstraints}
&\tilde{\Sigma}_3\lrcorner\,G~=~0\,,&&\tilde{\Sigma}_4\lrcorner\,G~=~0\,,&&\Sigma_5\lrcorner\,G~=~-3\,\Sigma_0\,\d A~.&
\end{align}
Splitting the flux $G$ into parts proportional and perpendicular to $v$, $G=F+v\w H$, and decomposing $F$ and $H$ under SU$(3)$
\begin{align}
\label{eq:Gdecomposition}
&F& &=& &A_1\,J\w J\,+\,A_2\w J\,+\,\overline{B}\w\Omega\,+\,B\w\overline{\Omega};&\\
&H& &=& &\overline{C}_1\,\Omega\,+\,C_1\overline{\Omega}\,+\,C_2\w J\,+\,C_3&\nonumber
\end{align}
one finds that $B=0$ and $C_1=0$.\footnote{$A_1$ is a real and $C_1$ a complex scalar, respectively. With respect to the almost complex structure defined by $J$ and $\Omega$ $A_2$ is primitive and $(1,1)$, $B$ is $(1,0)$, $C_2$ a real one-form, and $C_3$ is $(2,1)+(1,2)$ and primitive.} Hence $F$ is $(2,2)$ and $H$ is $(2,1)+(1,2)$. Furthermore the exterior derivative of the warp factor is determined by $A_1$ and $C_2$
\begin{equation}
\label{eq:GWarping}
\d A~=~2\,A_1\,v\,+\,\frac{2}{3}\,C_2\lrcorner\,J~=~a_1 \,v\,+\,a_2~.
\end{equation}
Plugging (\ref{eq:SU3TorsionClasses}) and (\ref{eq:Gdecomposition}) into (\ref{eq:DiffSUSYcond}) one obtains all SUSY conditions in terms of torsion classes
\begin{align}
\label{eq:TorsionConditions}
&R\,=\,0\,,&&V_1\,=\,T_1\,=\,0\,,&&W_0\,=\,2\,a_2\,,&\\
&E\,=\,\text{Re}E\,=\,-6\,A_1\,=\,-3\,a_1\,,&&W_1\,=\,\frac{8}{3}\,C_1\,=\,0\,=\,R\,,&&V_2\,=\,B\,=\,0\,,&\nonumber\\
&W_4\,=\,-\frac{1}{2}\,W_0\,=\,-\frac{2}{3}\,C_2\lrcorner\,J\,,&&W_2\,=\,S\,=\,0\,,&&T_2\,=\,-2\,A_2\,,&\nonumber\\
&\text{Re}W_5\,=\,-\,C_2\lrcorner\,J\,,&&\text{Im}W_5\,=\,-\,C_2\,,&&C_3\,=\,\frac{1}{2}\,v\lrcorner\,\ast W_3~.&\nonumber
\end{align}
This will turn out to be useful in our discussion of the scalar potential.
\section{Is a BPS-like potential possible?}
\label{sec:BPSPot?}
We are now ready to discuss whether a BPS-like potential is possible for compactifications of M-theory to four dimensions on SU$(3)$ structure manifolds. This would then be the case if the scalar potential (\ref{eq:potential0}) could be written as a sum of perfect squares containing the supersymmetry conditions (\ref{eq:DiffSUSYcond}). We will first consider the bulk potential and turn afterwards to the boundary contributions.
\subsection{Bulk potential}
Since we know from section \ref{sec:Supersymmetry conditions} that a supersymmetric vacuum must have $\mu=w_0=0$ we focus on these settings. This means that $\hat{R}^{(4)}$ and the both terms containing $\mu$ vanish in (\ref{eq:pot0bulk}) and we can start with
\begin{equation}
\label{eq:mu0potential}
V_0~=~-\,\frac{1}{2\,\kappa^2}\,\int\limits_{M}\text{dvol}_7{\,e^{4A}\,(R\,-\,8\nabla^2 A\,-\,20\,\d A^2\,-\,2|G|^2)}~,
\end{equation}
Comparing the formula for $R$ (\ref{eq:Rbuilding}) with (\ref{eq:DiffSUSYcond}) we see that in order to possibly match the differential supersymmetry conditions we have to insert the right powers of $e^A$ into the exterior derivatives of $v$, $J$, and $\Omega$, respectively. This will obviously lead to terms linear in $\d A$. Defining
\begin{align}
&\d\tilde{v}~=~e^{-2A}\d(e^{2A}\,v)\,,&&\d\tilde{\Omega}~=~e^{-3A}\d(e^{3A}\,\Omega)\,,&\\
&\d\tilde{J}~=~e^{-4A}\d(e^{4A}\,J)\,,&&\d\tilde{J}^2~=~e^{-2A}\d(e^{2A}\,J^2)\,,&\nonumber
\end{align}
we find in particular
\begin{align}
&\d\p(\d v,\d\Omega,\d J)&&=&&\d\p(\d\tilde{v},\d\tilde{\Omega},\d\tilde{J})\,-\,6\,\d A\w v\w J\,-\,3\,\d A\w\text{Re}\Omega~,
&\\
&\d\ps(\d v,\d\Omega,\d J^2)&&=&&\d\ps(\d\tilde{v},\d\tilde{\Omega},\d \tilde{J}^2)\,-\,\d A\w J^2\,-\,5\,\d A\w v\w\text{Im}\Omega
\,.&\nonumber
\end{align}
Additional linear terms will come from the derivative piece of $R$ in (\ref{eq:scalcurv}) after a partial integration. In order not to be bothered with boundary terms, we switch to the upstairs picture. We can then write\footnote{Here we used $(\d J\lrc v)\lrc\Re\Omega=-\Re\,\d\Omega\lrc (v\w J)$.}
\begin{align}
\label{eq:Rintermediate}
&\int\limits_{M}\text{dvol}_7\,e^{4A}\,\big\lbrace R\,-\,8\nabla^2 A\,-\,20\,\d A^2\big\rbrace\\
&~=~\int\limits_{M}\text{dvol}_7\,e^{4A}\,\Big\lbrace\frac{1}{2}\,\big|\d\tilde\ps\lrcorner\ps\big|^2\,-\,\frac{1}{2}\,\big|\d\tilde\p\big|^2\,-\,\frac{1}{2}\,\big|\d\tilde{\ps}\big|^2\,+\,\frac{1}{8}\,\big|\d\tilde{\p}\lrcorner\ps\big|^2\nonumber\\
&-\,18\,\big|\d A\lrcorner v\big|^2\,+\,3\,\text{Re}\,\d\tilde{\Omega}\lrcorner\big(\d A\w v\w J\big)\,-\,3\big(\d A\lrcorner v\big)\,\d\tilde{J}\lrcorner\text{Re}\Omega\,+\,6\,\d\tilde{v}\lrcorner\big(\d A\w v\big)\nonumber\\
&+\,\frac{3}{2}\,\big(\d A\lrcorner v\big)\,\text{Re}\big[\d\tilde{\Omega}\lrcorner\big(v\w\overline{\Omega}\big)\big]\,-\,3\big(\d A\w\d\tilde{v}\big)\lrcorner\text{Im}\Omega\Big\rbrace~.\nonumber
\end{align}
Here $\d\tilde{\p}$ and $\d\tilde{\ps}$ are shorthand notations for $\d\p(\d\tilde{v},\d\tilde{\Omega},\d\tilde{J})$ and $\d\ps(\d\tilde{v},\d\tilde{\Omega},\d \tilde{J}^2)$, respectively. Clearly, all but the first four terms of this expression vanish at most linear if the conditions (\ref{eq:DiffSUSYcond}) are imposed. If it is not possible to cancel them, then a BPS-like form of $V$ will not be available.

In order to see if such a cancellation happens we have to involve the flux in our discussion. The first four terms of (\ref{eq:Rintermediate}) contain exterior derivatives $\d\tilde{J}$ and $\d\tilde{J^2}$. If SUSY is to be maintained after the compactification these should be proportional to the flux $G$. Inserting $G$ will also lead to contributions that do not vanish quadratically under SUSY. Schematically these contributions will look like
\begin{equation}
|\d J\lrc U\,+\, V|^2~=~|(\d J\,-\,2\ast G)\lrc U\,+\,V|^2\,-\,8\,|\ast G\lrc U|^2\,-\,4\,(\ast\,G\lrc U)\lrc(\d J\lrc U +V)
\end{equation}
and could eventually cancel the terms in (\ref{eq:Rintermediate}). But as it turns out a direct insertion is very cumbersome and not very enlightening.

Instead, we split the derivatives of $v$, $J$, and $\Omega$ in their parts proportional and perpendicular to $v$
\begin{align}
\label{eq:dsplitting}
&\d\tilde{v}&&=&&\d\tilde{v}_{\perp}\,+\,v\w(\d\tilde{v}\lrc v)&&\d\tilde{\Omega}&&=&&\d\tilde{\Omega}_{\perp}\,+\,v\w(\d\tilde{\Omega}\lrcorner v)&\\
&\d\tilde{J}&&=&&\d\tilde{J}_{\perp}\,+\,v\w(\d\tilde{J}\lrcorner v)&&\d\tilde{J^2}&&=&&\d\tilde{J^2}_{\perp}\,+\,v\w(\d\tilde{J^2}\lrcorner v)~.&\nonumber
\end{align}
In particular, it is the fact that $\d\tilde{J}_{\perp}^2$ will vanish for supersymmetric vacua due to $\d\tilde{J}^2=-4v\w G$ that simplifies the calculation. The square terms in (\ref{eq:Rintermediate}) can be brought to the form
\begin{subequations}
\label{eq:splitting}
\begin{align}
&-\frac{1}{2}\big|\d\tilde{\p}\big|^2\,=\,-\frac{1}{2}\big|\d\tilde{\p}_{\perp}\big|^2-(\text{Re}\,\d\tilde{\Omega}\lrcorner v)\lrc\Big[\frac{1}{2}\text{Re}\,\d\tilde{\Omega}\lrcorner v+(\d\tilde{v}\lrc v)\w J-\d\tilde{J}_{\perp}\Big]\,,\label{eq:splitting1}\\
&\nonumber\\
&-\frac{1}{2}\big|\d\tilde{\ps}\big|^2=-\,\frac{1}{2}\big|\d\tilde{\ps}_{\perp}\big|^2+\text{Im} \d\tilde{\Omega}_{\perp}\lrcorner\big[J\w(\d\tilde{J}\lrcorner v)\big]-3(\d A\lrc v)\,\Im \d\tilde{\Omega}_{\perp}\lrc J^2\label{eq:splitting2}\\
&~~~~~~~~~~~~~~~~-\frac{1}{8}\big|\d\tilde{J}^2\lrcorner v\big|^2-\Re\,\d\tilde{\Omega}\lrc(v\w J\w (\d\tilde{v}\lrc v))\,,\nonumber\\
&\nonumber\\
&\frac{1}{2}\big|\d\tilde{\ps}\lrcorner\ps\big|^2\,=\,\frac{1}{2}\big|\d\tilde{\ps}_{\perp}\lrcorner\ps\big|^2+\frac{1}{8}\Big[\d\tilde{J}^2\lrcorner(v\w J^2)\Big]\lrc\Big[\frac{1}{4}\d\tilde{J}^2\lrcorner(v\w J^2)-\big(\text{Im} \d\tilde\Omega_{\perp}\lrcorner J^2\big)\Big]\label{eq:splitting3}\\
&~+\Big[\Re\,\d\tilde{\Omega}\lrc(v\w J)\Big]\lrc\Big[\frac{1}{2}\Re\,\d\tilde{\Omega}\lrc(v\w J)-\d\tilde{v}_{\perp}\lrcorner\text{Im}\Omega-\frac{1}{4}\d\tilde{J}^2_{\perp}\lrcorner J^2+\frac{1}{2}\text{Re}(\d\tilde{\Omega}_{\perp}\lrcorner\overline{\Omega})+2\d\tilde{v}\lrc v\Big]\,,\nonumber\\
&\nonumber\\
&\frac{1}{8}\big|\d\tilde{\p}\lrcorner\ps\big|^2=\frac{1}{8}\big|\d\tilde{\p}_{\perp}\lrcorner\ps\big|^2+\frac{1}{8}\Im[ \d\tilde{\Omega}\lrc(v\w\bar{\Omega})]\hskip-2pt\times\hskip-2pt\Big[\frac{1}{4}\Im[ \d\tilde{\Omega}\lrc(v\w\bar{\Omega})]-2\d\tilde{v}_{\perp}\lrc J-\Re\,\d\tilde{\Omega}_{\perp}\lrc J^2\Big]\,.
\end{align}
\end{subequations}
Note that in each expression there is one term including $\d\tilde{\p}_{\perp}=\d\p(\d\tilde{v},\d\tilde{\Omega}_{\perp},\d\tilde{J}_{\perp})$ or $\d\tilde{\ps}_{\perp}~=~\d\ps(\d\tilde{v},\d\tilde{\Omega}_{\perp},\d \tilde{J}^2_{\perp})$. We find that in the combination of these
\begin{align}
\label{eq:splitsquares}
&\frac{1}{2}\,\big|\d\tilde{\ps}_{\perp}\lrcorner\ps\big|^2+\frac{1}{8}\,\big|\d\tilde{\p}_{\perp}\lrcorner\ps\big|^2-\frac{1}{2}\,\big|\d\tilde{\p}_{\perp}\big|^2-\frac{1}{2}\,\big|\d\tilde{\ps}_{\perp}\big|^2~=~\\
&\hskip-2pt-\frac{1}{2}\,|\d \tilde{J}_{\perp}|^2-\frac{1}{2}\,|\d\tilde{\Omega}_{\perp}|^2-\frac{1}{8}\,|\d \tilde{J}_{\perp}^2|^2+\frac{1}{2}\,\Big|\frac{1}{4}\,\d\tilde{J}^2_{\perp}\lrcorner J^2-\frac{1}{2}\,\text{Re}(\d\tilde{\Omega}_{\perp}\lrcorner\overline{\Omega})-\d\tilde{v}\lrc v\Big|^2+\frac{1}{8}\,\big|\d\tilde{\Omega}_{\perp}\lrcorner J^2\big|^2\nonumber \\
&\hskip-2pt-\frac{1}{2}\,\big|\d\tilde{v}\lrc v\big|^2-\frac{1}{2}\,\big|\d\tilde{v}_{\perp}\big|^2-2\,(\d\tilde{v}_{\perp}\lrc\Im\,\Omega)\lrc (\d\tilde{v}\lrc v)+\frac{1}{6}(\Re\,\d\tilde{\Omega}_{\perp}\w \d\tilde{v}_{\perp})\lrc J^3-6\d\tilde{v}\lrcorner (\d A\w v)\nonumber
\end{align}
only the last term and $|\d \tilde{J}_{\perp}|^2$ do not vanish quadratically when supersymmetry is imposed. Note that in order to obtain this form we used the identities
\begin{align}
&\Im\,\d\tilde{\Omega}_{\perp}\lrc\big[(\d\tilde{v}_{\perp}\lrc\Im\Omega)\w\Im\Omega\big]+\frac{1}{2}(\d\tilde{v}_{\perp}\lrc J)(\Re\,\d\tilde{\Omega}_{\perp}\lrc J^2)-\Re\,\d\tilde{\Omega}_{\perp}\lrc(\d\tilde{v}_{\perp}\w J)~=\\
&~~~~~~~~~~~~~~~~~~=~\frac{1}{6}(\Re\,\d\tilde{\Omega}_{\perp}\w \d\tilde{v}_{\perp})\lrc J^3~,\nonumber\\
&\big|\d v\w\Im\Omega\big|^2\,-\,\big|\d v\lrc\Im\Omega\big|^2~=~2\big|\d v\lrc v\big|^2\,,~~~\big|\d v\w J\big|^2\,-\,\big|\d v\lrc J\big|^2~=~\big|\d v_\perp\big|^2\,+2\,\big|\d v\lrc v\big|^2~.\nonumber
\end{align}
Since in the end we will integrate over the whole expression we can even get a further simplification using partial integration and the fact that $2\,(\d\tilde{v}_{\perp}\lrc\Im\,\Omega)\lrc (\d\tilde{v}\lrc v)=(\d\tilde{v}\w\d\tilde{v})\lrc(v\w\Im\Omega)$
\begin{equation}
\label{eq:integralidentity}
\int\limits_{M}\,e^{4A}\,\Big\lbrace\frac{1}{6}(\Re\,\d\tilde{\Omega}_{\perp}\w \d\tilde{v}_{\perp})\lrc J^3\,-\,(\d\tilde{v}\w\d\tilde{v})\lrc(v\w\Im\Omega)\Big\rbrace\,=\,3\int\limits_{M}\,e^{4A}\,(\d A\w\d \tilde{v})\lrc\Im\Omega~.
\end{equation}
This will cancel exactly against the last term appearing in (\ref{eq:Rintermediate}). We thus conclude that we can neglect all terms including $\d\tilde{v}_{\perp}$ in (\ref{eq:splitsquares}) except for $-\frac{1}{2}|\d\tilde{v}_{\perp}|^2$ as long as we also neglect the term $-3(\d A\w\d \tilde{v})\lrc\Im\Omega$ from (\ref{eq:Rintermediate}). We also see that the last term of (\ref{eq:splitsquares}) will cancel against a term in (\ref{eq:Rintermediate}).

Examining the rest of (\ref{eq:splitting1}) - (\ref{eq:splitting3}) we find only six more terms that do not vanish quadratically under supersymmetry
\begin{align}
\label{eq:sixnonzero}
&\text{Re}\,\d\tilde{\Omega}\lrcorner\big( v\w\d\tilde{J}_{\perp}\big)\,,& &\phantom{m}\big[J\w(\d\tilde{J}\lrcorner v)\big]\lrc\text{Im} \d\tilde{\Omega}_{\perp}\,,& &-\frac{1}{8}\big|\d\tilde{J}^2\lrcorner v\big|^2\,,&\\
&\frac{1}{32}\big|\d\tilde{J}^2\lrcorner(v\w J^2)\big|^2\,,& &-\frac{1}{8}\Big[\d\tilde{J}^2\lrcorner(v\w J^2)\Big]\lrc\big(\text{Im} \d\tilde\Omega_{\perp}\lrcorner J^2\big)\,,& &-3(\d  A\lrc v)\Im\d\tilde{\Omega}\lrc J^2\,.&\nonumber
\end{align}
This means that due to the split (\ref{eq:dsplitting}) we have reduced the number of squares that do not vanish under SUSY, and which thus should be combined with $G$ flux, to three. To check whether from these terms can come contributions that cancel the other linearly vanishing expressions, or if some of them cancel themselves, we will in the end expand all expressions in terms of the SU$(3)$ torsion classes (\ref{eq:SU3TorsionClasses}) and use the SUSY conditions in the form (\ref{eq:TorsionConditions}).

But before we do so it is important to note that we have not yet used the Bianchi identity of the four-form flux $G$. In \cite{Lust:2008zd,Held:2010az} the use of the Bianchi identity was a keystone in order to obtain a BPS-like potential. We follow these references and implement the Bianchi identity by a partial integration
\begin{align}
\label{eq:BIBoundary}
&\int\limits_{M}\text{dvol}_7\,e^{4A}\,\big\lbrace -\frac{1}{2}\,|\d \tilde{J}_{\perp}|^2-2\,\big|G\big|^2 \big\rbrace~=~\int\limits_{M}\text{dvol}_7\,e^{4A}\,\big\lbrace -\frac{1}{2}\,|\d \tilde{J}|^2-2\,\big|G\big|^2+\frac{1}{2}\,|\d\tilde{J}\lrc v|^2 \big\rbrace~= \nonumber\\
&\int\limits_{M}\text{dvol}_7\,e^{4A}\,\big\lbrace -\frac{1}{2}\,|\d \tilde{J}-2\ast G|^2+\,\d G\lrc(v\w J^2)+\frac{1}{2}\,|\d\tilde{J}\lrc v|^2 \big\rbrace~=\\
&\int\limits_{M}\text{dvol}_7\,e^{4A}\,\big\lbrace -\frac{1}{2}\,|\d \tilde{J}-2\ast G|^2+\frac{1}{2}\,|\d\tilde{J}\lrc v|^2 \big\rbrace\,-\,\frac{1}{4\pi}\,\Big(\frac{\kappa}{4\pi}\Big)^{2/3}\sum\limits_{p=1,2}\,\int\limits_{B_{6,p}} e^{4A}\,J\w K^{(p)},\nonumber
\end{align}
where we used (\ref{eq:GBI}) in the last step. We conclude that we have to include $\frac{1}{2}\,|\d\tilde{J}\lrc v|^2$ into our analysis of the bulk action in order to take the Bianchi identity of $G$ into account. Furthermore, we get an additional contribution to the boundary action.

All things considered in order to obtain a BPS-like form of the potential the sum of $\frac{1}{2}\,|\d\tilde{J}\lrc v|^2$, the remaining linear terms of (\ref{eq:Rintermediate}), and the six terms of (\ref{eq:sixnonzero})
\begin{align}
\label{eq:linearpieces}
L\,=\,&\frac{1}{2}\,|\d\tilde{J}\lrc v|^2-\frac{1}{8}\big|\d\tilde{J}^2\lrcorner v\big|^2+\frac{1}{32}\big|\d\tilde{J}^2\lrcorner(v\w J^2)\big|^2-\frac{1}{8}\Big[\d\tilde{J}^2\lrcorner(v\w J^2)\Big]\lrc\big(\text{Im} \d\tilde\Omega_{\perp}\lrcorner J^2\big)\\
&-18\big|\d A\lrcorner v\big|^2+\big[J\w(\d\tilde{J}\lrcorner v)\big]\lrc\text{Im} \d\tilde{\Omega}_{\perp}+\text{Re}\,\d\tilde{\Omega}\lrcorner\big( v\w\d\tilde{J}_{\perp}\big)+3\,\text{Re}\,\d\tilde{\Omega}\lrcorner\big(\d A\w v\w J\big)\nonumber\\
&-3\big(\d A\lrcorner v\big)\,\d\tilde{J}\lrcorner\text{Re}\Omega-3\big(\d A\lrcorner v\big)\,\text{Im}\,\d\tilde{\Omega}\lrcorner J^2+\frac{3}{2}\big(\d A\lrcorner v\big)\,\text{Re}\big[\d\tilde{\Omega}\lrcorner\big(v\w\overline{\Omega}\big)\big]\nonumber
\end{align}
has to vanish quadratically for a supersymmetric setting. Inserting the expansion (\ref{eq:SU3TorsionClasses}) and reordering terms we get
\begin{align}
L\,=\,&\frac{3}{2}\big|(\d\tilde{J}\lrc(v\w J)-6\,\d A\lrc v\big|^2+(\Re\d\tilde{\Omega}\lrc v)\lrc\big[\d\tilde{J}-3\,\d A\w J\big]&\label{eq:noBPS2}\\
&+\Im\d\tilde{\Omega}\lrc(J\w (\d\tilde{J}\lrc v))-24\,\Re E\Im\, W_1-90\,(\d A\lrc v)\Im\, W_1&\nonumber\\
=\,&6\,|\Re E+3\,\d A\lrc v|^2-10\,\Im \,W_1(\Re E+3\,\d A\lrc v)-8\,\Re \,V_2\lrc (2\,\Re \,W_5+ W_4+4\,\d A)\nonumber\\
&+6\,\Im \,E\,\Re \,W_1+6\,(\d A\lrc v)\Im \,W_1+T_2\lrc\Im \,W_2+\Re\, S\lrc W_3~.\label{eq:noBPS}
\end{align}
Using the relations (\ref{eq:TorsionConditions}) we see that all except of the last three terms of (\ref{eq:noBPS}) will indeed go to zero quadratically under SUSY. The last three terms vanishes linearly since $\d A\lrc v$, $T_2$, and $W_3$ are non-zero generically. However, we can rewrite $(\Re E\,\Im W_1)$ using partial integration and the fact that $\Im\,\d W_1\lrc v=0$ under SUSY\footnote{Here we used $\Re E\,\d\text{vol}_7=\frac{1}{2}\d J\lrc(v\w J)\,\d\text{vol}_7=\frac{1}{4}\d J\w J\w J=\frac{1}{12}\d J^3$.}
\begin{equation}
\int\limits_{M}\text{dvol}_7\,e^{4A}\,\Re E\,\Im W_1\,=\,-2\,\int\limits_{M}\text{dvol}_7\,e^{4A}\,(\d A\lrc v)\Im W_1~.
\end{equation}
Another partial integration shows that $(\d A\lrc v)\Im W_1$ will vanish quadratically under the integral, since
\begin{equation}
\int\limits_{M}\text{dvol}_7\,e^{4A}\,(\d A\lrc v)\Im W_1\,=\,-\frac{1}{4}\int\limits_{M}\text{dvol}_7\,e^{4A}\big[(\partial^m v_m)\,\Im W_1\,+\Im\,\d W_1\lrc v\big]~.
\end{equation}
The second term on the right hand side gives zero, and therefore one obtains
\begin{equation}
\int\limits_{M}\text{dvol}_7\,e^{4A}\,\big[(\d A\lrc v)\,+\,\frac{1}{4}\,(\partial^m v_m)\big]\Im W_1~=~0~.
\end{equation}
$\Im W_1$ is an arbitrary function (not depending on the direction of $v$) and can be viewed as a test function. Thus $e^{4A}\,\big[(\d A\lrc v)\,+\,\frac{1}{4}\,(\partial^m v_m)\big]$ integrates to zero. But under SUSY (\ref{eq:inSUSYcond}) gives $\partial^m v_m=\nabla^m v_m=7(\d A\lrc v)$ which means that $e^{4A}(\d A\lrc v)$ will also integrate to zero when SUSY is imposed. Since $\Im W_1$ is zero in this case, too, we see that $\smash{\int\limits_{M}\text{dvol}_7\,e^{4A}\,(\d A\lrc v)\Im W_1}$ will vanish quadratically in a supersymmetric setting.

Unfortunately, we did not see how one could argue in a similar way for the last two terms of (\ref{eq:noBPS}). Thus we conclude with the surprising result that M-theory compactified on a general seven dimensional SU$(3)$ structure manifold does not admit a BPS-like scalar potential, since in general the torsion classes $T_2$ and $W_3$ do not vanish. 

Gathering all terms together the bulk potential reads
\begin{align}
V_0\,=\,&\frac{1}{4\kappa^2}\int\limits_{M}\text{dvol}_7\,e^{4A}\Big\lbrace\,|\d \tilde{J}-2\ast G|^2+|\d\tilde{\Omega}_{\perp}|^2+\frac{1}{4}\,|\d \tilde{J}_{\perp}^2|^2-\frac{1}{4}\,\big|\d\tilde{\Omega}_{\perp}\lrcorner J^2\big|^2 \\
&-\Big|\frac{1}{4}\,\d\tilde{J}^2_{\perp}\lrcorner J^2-\frac{1}{2}\,\text{Re}(\d\tilde{\Omega}_{\perp}\lrcorner\overline{\Omega})-\d\tilde{v}\lrc v-\Re\,\d\tilde{\Omega}\lrc(v\w J)\Big|^2+\big|\d\tilde{v}\lrc v\big|^2+\big|\d\tilde{v}_{\perp}\big|^2\nonumber\\
&+\big|\text{Re}\,\d\tilde{\Omega}\lrcorner v\big|^2+\frac{1}{4}\,\Im[ \d\tilde{\Omega}\lrc(v\w\bar{\Omega})]\times\Big[\frac{1}{4}\,\Im[ \d\tilde{\Omega}\lrc(v\w\bar{\Omega})]-2\,\d\tilde{v}_{\perp}\lrc J-\Re\,\d\tilde{\Omega}_{\perp}\lrc J^2\Big]\nonumber\\
&+2\,\Re\,\d\tilde{\Omega}\lrc\big(v\w J\w(\d\tilde{v}_{\perp}\lrcorner\text{Im}\Omega)\big)-3\big|\d\tilde{J}\lrc (v\w J)-6\d A\lrc v\big|^2+7\,(\d A\lrc v)\Im\d\tilde{\Omega}_\perp\lrc J^2\nonumber\\
&-2(\Re\,\d\tilde{\Omega}\lrc v)\lrc\big[\d\tilde{J}_\perp-3\d A\w J\big]-2\Im\d\tilde{\Omega}_\perp\lrc(J\w (\d\tilde{J}\lrc v))\Big\rbrace~.\nonumber
\end{align}
One should notice here that it are the last two terms that spoil the BPS-like form
\begin{align}
\label{eq:nonBPSpot}
&V_{\text{no-BPS}}\,=\,-\frac{1}{2\kappa^2}\int\limits_{M}\text{dvol}_7\,e^{4A}\Big\lbrace(\Re\,\d\tilde{\Omega}\lrc v)\lrc\big[\d\tilde{J}_\perp-3\d A\w J\big]+\Im\d\tilde{\Omega}_\perp\lrc(J\w (\d\tilde{J}\lrc v))\Big\rbrace\nonumber\\
&~~\,=\,-\frac{1}{2\kappa^2}\int\limits_{M}\text{dvol}_7\,e^{4A}\Big\lbrace\,24\,\Im\,W_1(\d A\lrc v)+\Re\,S\lrc W_3+\Im\,W_2\lrc T_2+6\,\Im\,E\,\Re\,W_1\\
&~~~~~~+14\,\Im\,W_1(\Re\,E+3\,\d A\lrc v)-8\,\Re\,V_2\lrc(2\,\Re\,W_5+W_4+4\d A)\Big\rbrace~.\nonumber
\end{align}
However, for e.g. $S=0=T_2$ also this part of the potential reduces to a BPS-like form
\begin{align}
\label{eq:nonBPSpotS0}
&V_{\text{no-BPS}}\,\mathop{=}_{T_2=0}^{S=0}\,-\,\frac{1}{2\kappa^2}\int\limits_{M}\text{dvol}_7\,e^{4A}\Big\lbrace \frac{1}{2}\big[\d\tilde{J}_\perp\lrc J-6\,a_2-\frac{1}{2}\Re(\d\tilde{\Omega}_\perp\lrc\bar{\Omega})\big]\lrc\big[\Re\tilde{\Omega}\lrc (v\w J)\big] \\
&~~~~~~~~~~~~~~~~~~~~~~~~~~~~~~~~~+\frac{1}{16}\Im[\d\tilde{\Omega}\lrc(v\w \bar{\Omega})]\,\Re\,\d\tilde{\Omega}_\perp\lrc J^2+\frac{19}{6}(\d A\lrc v)\Im\,\d\tilde{\Omega}_\perp\lrc J^2\Big\rbrace~.\nonumber
\end{align}
So we wee that it is in general not possible to bring the bulk part of the potential to a BPS-like form. But by setting the terms containing $T_2$ and $W_3$ to zero such a form can be reached.
\subsection{Boundary potential}
\label{sec:BouPot}
The boundary potential receives contributions from two sources. Besides of (\ref{eq:pot0boundary}) one also has to include the boundary piece obtained by integration over the Bianchi identity in (\ref{eq:BIBoundary}). Before combining the two pieces one has to make sure that both are given in terms of the same metric $g'_{10}$ that we introduced in section \ref{sec:ScalPotAndEOM}. However, going from $g_{10}$ to $g'_{10}$ does not lead to new contributions from $K^{(p)}$. Thus the boundary potential is given by
\begin{align}
&V_b~=~\frac{1}{8\pi\kappa^2}\Big(\frac{\kappa}{4\pi}\Big)^{2/3}\sum\limits_{p=1,2}\,\int\limits_{B_{6,p}}\text{dvol}'_{6,p}\,e^{4A'-2\phi}\,\Big\lbrace -\frac{1}{24}\,\big|e^{-2A'}\,\hat{R}^{(4)}-12|\d A'|^2\big|^2\\
&~~~~+\Big(\text{Tr}|\calf^{(6)}_{p}\lrc J|^2\,+\,2\,\text{Tr}|(\calf^{(6)}_{p})^{2,0}|^2\Big)-\,\frac{1}{2}\Big(\text{Tr}|R^{(6)'}_{p\,+}\lrc J|^2\,+\,2\,\text{Tr}|(R^{(6)'}_{p\,+})^{2,0}|^2\Big)\nonumber\\
&~~~~-4\,e^{-2A'}\,(\nabla_i\nabla_j e^{A'})(\nabla^i\nabla^j e^{A'})-8\,\big|\d A'\lrc H\big|^2\Big\rbrace~.\nonumber
\end{align}
Since for a supersymmetric vacuum we have to restrict to Minkowski space $\hat{R}^{(4)}$ will vanish. Also the terms containing $\calf_p$ will vanish by the SUSY conditions (\ref{eq:Fcond}). Let us consider next the $\d A'$ terms. From (\ref{eq:DiffSUSYv}), (\ref{eq:GWarping}), and the relation $A'=A+\frac{1}{2}\sigma$ we know that
\begin{equation}
-2\,\d A'\,+\,\d\sigma_{\perp}\,+\,\d v\lrc v~=~0~,
\end{equation}
where $\d\sigma_{\perp}$ denotes as usual the part of $\d\sigma$ that is perpendicular to $v$. Thus in order to obtain $\d A'=0$ for a SUSY vacuum the identity $\d\sigma_{\perp}=-W_0$ should hold. Since we did not specify $\sigma$ yet, we choose it in such a way that the above equation holds. In section~\ref{sec:10dlimit} we will see a justification for this choice.

The last terms to consider are the $R^{(6)'}_{p\,+}$ terms. These vanish if $R^{(6)'}_{p\,+}$ is $(1,1)$ and primitive. For the heterotic string this can be shown by using the integrability condition
\begin{equation}
\Big[\nabla^{\text{het}}_{i},\nabla^{\text{het}}_{j}\Big]\eta_{\text{het}}~=~\frac{1}{4}\,R_{klij}\gamma^{kl}\,\eta_{\text{het}}~,
\end{equation}
where 'het' denotes that the various objects belong to the heterotic string (see e.g. \cite{Sen:1986mg,LopesCardoso:2003af,Held:2010az}). We thus trace back the problem to the heterotic setting. In order to do so
one has to determine how the seven dimensional covariant derivative $\nabla_m$ which appears in (\ref{eq:inSUSYcond}) is related to its six dimensional counterpart at the boundary. This is an quite easy task in type IIA supergravity, where the geometry close to the boundary can be chosen to be independent of the extra dimension (see e.g. \cite{Becker:2007zj}). In Horava-Witten theory this gets changed by the non-vanishing of the four-form flux Bianchi identity. However, the modifications appear only at order $\kappa^{2/3}$. Since the boundary terms are already of order $\kappa^{2/3}$ one can consistently neglect the corrections and work in the type IIA setting
\begin{equation}
\d s_7^2~=~e^{-\sigma}(\d s'_6)^2\,+\,e^{2\sigma}\d x^{11}~
\end{equation}
with $\sigma$ and $g'_6$ independent of $x^{11}$. A calculation along the lines of \cite{Becker:2007zj} shows then that
\begin{equation}
\nabla^{(6)'}_{-\,i}\big(e^{\frac{\sigma}{4}}\eta_+\big)~=~\nabla^{(6)'}_i\big(e^{\frac{\sigma}{4}}\eta_+\big)\,-\,\frac{1}{4}G_{11\,ijk}\,\gamma^{jk}\,\big(e^{\frac{\sigma}{4}}\eta_+\big)~=~\mathcal{O}(\kappa^{2/3})~.
\end{equation}
Then, on each boundary
\begin{equation}
\Big[\nabla^{(6)'}_{-\,i},\nabla^{(6)'}_{-\,j}\Big]\,\big(e^{\frac{\sigma}{4}}\eta_+\big)~=~e^{\frac{\sigma}{4}}\,\Big[\nabla^{(6)'}_{-\,i},\nabla^{(6)'}_{-\,j}\Big]\,\eta_+~=~\frac{1}{4}e^{\frac{\sigma}{4}}\,R^{(6)'}_{-\,mnij}\,\gamma^{mn}\,\eta_+~=~\mathcal{O}(\kappa^{2/3})~,
\end{equation}
which is precisely the condition one obtains for the heterotic string. From this it follows that $R^{(6)'}_{p\,+}$ is $(1,1)$ and primitive up to $\mathcal{O}(\kappa^{2/3})$ which is sufficient for our analysis as the boundary potential is already of order $\kappa^{2/3}$.
\vskip12pt
We conclude that the boundary potential can be rewritten in a BPS-like form, although this is not possible for the bulk potential. The fact that a BPS-like form of the potential is not available in general does of course not mean that our ansatz is inconsistent. It merely states that in addition to the Bianchi identity and the SUSY conditions the ansatz has to satisfy further restrictions that come from the variation of (\ref{eq:nonBPSpot}) in order to ensure the equations of motion. On the other hand (\ref{eq:nonBPSpotS0}) tells us how to restrict our compactification ansatz if we wish to get a BPS-like scalar potential. For example if one chooses a manifold for which $T_2$, and $\Re\,S$ vanish identically (and not only if SUSY is imposed) then the whole action can be written in terms of squares of the supersymmetry conditions. However it would be nice to see whether our findings give the correct results when restricted to well known geometrical settings. This will be discussed in the next section.
\section{Limiting cases}
\label{sec:limits}
In order to strengthen our results we will show that they reduce correctly in the three cases of G$_2$ holonomy, six dimensional SU$(3)$ holonomy, and the heterotic limit.
\subsection{G$_2$ holonomy}
It is well known that compactifications on manifolds with G$_2$ holonomy do not allow four-form flux $G$. Hence they are not viable for Horava-Witten theory where $G$ is necessarily not zero. Nevertheless, one can ask whether our formulas behave in the right way in this limit although we know that they will not give a solution for heterotic M-theory. In particular, we expect the curvature scalar $R$ to be zero for a G$_2$ holonomy manifold $M$. Furthermore, once the SUSY conditions are satisfied also the $G$ flux should be set to zero and the warp factor $A$ should be constant.

G$_2$ holonomy is specified by the conditions
\begin{equation}
\label{eq:G2Holonomy}
\d\p~=~0~~~~~~~~~~~\text{and}~~~~~~~~~~~\d\ps~=~0~.
\end{equation}
Applying these conditions to the decomposition (\ref{eq:PhiPsiDecompositons}) of $\p$ and $\ps$ in SU$(3)$ structure forms one finds that for G$_2$ holonomy manifolds
\begin{align}
\label{eq:G2HolConditions}
&\Re W_1\,=\,\frac{2}{3}\,\Im E\,=\,-R\,,&&\Im W_1\,=\,\frac{2}{3}\,\Re E\,,&&T_1\,=\,-\Re W_2\,,&&T_1\,=\,\Im W_2\,,&\\
&\Re V_1\,=\,\frac{1}{2}\,\Im W_5\,,&&\Im V_1\,=\,-\frac{1}{2}\,\Re W_5\,,&&\Re S\,=\,W_3\,,&&J\lrc W_0\,=\,2\Im V_2\,,&\nonumber\\
&W_0~=~W_4\,\,+\,\,4\,\,\Re V_2~=&&\Re W_5\,\,+\,\,2\,\,\Re V_2~.&\nonumber
\end{align}
These conditions do clearly not imply that all SU$(3)$ torsion classes vanish. This means that although it is clear from (\ref{eq:G2Holonomy}) and (\ref{eq:scalcurv}) that $R=0$ it is a non-trivial consistency check for our results that the scalar curvature also vanishes in (\ref{eq:Rintermediate}) for a G$_2$ manifold.

In order to check the equations (\ref{eq:splitting}), too,  we split $\d\p$ and $\d\ps$ into parts proportional and perpendicular to $v$ and find the four conditions
\begin{align}
&\Re\,\d\tilde{\Omega}_{\perp}&&=&&3\,a_2\w\Re \Omega\,,&\nonumber\\
&\frac{1}{2}\,\d\tilde{J}^2_{\perp}&&=&&a_2\w J^2\,-\,\d\tilde{v}_\perp\w\Im\Omega\,,&\\
&\Im\,\d\tilde{\Omega}_\perp&&=&&\frac{1}{2}\,\d\tilde{J}^2\lrc v\,+\,(5\,a_2\,+\,\d\tilde{v}\lrc v)\w\Im \Omega\,-\,(\d A\lrc v)\,J^2\,,&\nonumber\\
&\d\tilde{J}_{\perp}&&=&&(6\,\d A\,+\,\d\tilde{v}\lrc v)\w J\,+\,\Re\d\tilde{\Omega}\lrc v\,-\,3\,(\d A\lrc v)\,\Re\Omega~.&\nonumber
\end{align}
Plugging these into (\ref{eq:splitting}) one can express the first line of (\ref{eq:Rintermediate}) solely in terms of $\d A\lrc v$
\begin{equation}
\frac{1}{2}\,\big|\d\tilde\ps\lrcorner\ps\big|^2\,-\,\frac{1}{2}\,\big|\d\tilde\p\big|^2\,-\,\frac{1}{2}\,\big|\d\tilde{\ps}\big|^2\,+\,\frac{1}{8}\,\big|\d\tilde{\p}\lrcorner\ps\big|^2~=~-6\,\d A\lrc v~,
\end{equation}
while the rest of (\ref{eq:Rintermediate}) reduces to $12|a_2|^2+18|\d A\lrc v|^2$. We thus see that
\begin{equation}
\int\limits_{M}\text{dvol}_7\,e^{4A}\,\big\lbrace R\,-\,8\nabla^2 A\,-\,20\,\d A^2\big\rbrace~=~12\,\int\limits_{M}\text{dvol}_7\,e^{4A}\,|\d A|^2~,
\end{equation}
which is just what one gets by setting $R=0$ and integrating by parts. This means that our formulas give indeed the right results in the G$_2$ holonomy limit.

Coming to the SUSY conditions one immediately sees, that the combination of the conditions (\ref{eq:TorsionConditions}) and (\ref{eq:G2HolConditions}) leads necessarily to the vanishing of all torsion classes and all flux components. Since this is what is expected for a G$_2$ holonomy compactifications, we see that also here our formulas provide the correct answer.
\subsection{SU$(3)$ holonomy}
\label{sec:SU3Hol}
Next we consider manifolds $M$ that obey
\begin{equation}
\d J_\perp~=~0\,~~~~\text{and}~~~~\d\Omega_\perp~=~0\,.
\end{equation}
This means that locally $M$ splits into a six dimensional manifold of SU$(3)$ holonomy (i.e. a Calabi-Yau three-fold) times a line in the direction of $v$. Globally however there can still be dependencies of $J$ and  $\Omega$ on $v$ and hence $\d J\neq 0\neq \d\Omega$. In terms of torsion classes this can be achieved by setting $W_i=0$ for $i=1,\ldots,5$. Since this will cancel all terms that do not vanish quadratically when SUSY is imposed from (\ref{eq:noBPS}), for this case a BPS-like form of the bulk potential is possible.

But before we come to the potential let us again check whether our formulas are consistent. We have now
\begin{equation}
\d\p_{CY}~=~\d v\w J\,+\,v\w(\Re\,\d\Omega\lrc v)~~~~\text{and}~~~~\d\ps_{CY}~=~\frac{1}{2}\,v\w(\d J^2\lrc v)\,+\,\d v\w\Im\Omega~.
\end{equation}
Plugging these directly into (\ref{eq:scalcurv}) we find
\begin{align}
\label{eq:CYLimit}
&\int\limits_{M}\text{dvol}_7\,e^{4A}\,R~=~\int\limits_{M}\text{dvol}_7\,e^{4A}\,\Big\lbrace\frac{1}{32}\big|\d J^2\lrc(v\w J^2)\big|^2-\frac{1}{8}\big|\d J^2\lrc v\big|^2-\frac{1}{2}\,\big|\Re\,\d\Omega\lrc v\big|^2\\
&~-\frac{1}{2}\,\big|\d v_\perp\big|^2+\big(\Re\,\d\Omega\lrc(v\w J)\big)\lrc\Big[\frac{1}{2}\Re\,\d\Omega\lrc(v\w J)-\Im\Omega\lrc\d v_{\perp}\Big]-2\,\Im\Omega\lrc\big((\d v\lrc v)\w \d v_\perp\big)\nonumber\\
&~+\frac{1}{8}\Im[ \d\Omega\lrc(v\w\bar{\Omega})]\times\Big[\frac{1}{4}\Im[ \d\Omega\lrc(v\w\bar{\Omega})]-2\d\tilde{v}_{\perp}\lrc J\Big]-4\,\Re\,\d\Omega\lrc(v\w a_2\w J)\nonumber\\
&~+4\,\Im\Omega\lrc(a_2\w \d v_\perp)+8\,\d v\lrc(a_2\w v)+(\d A\lrc v)\,\big(\d J^2\lrc(v\w J^2)\big)\Big\rbrace~.\nonumber
\end{align}
On the other hand, working with (\ref{eq:Rintermediate}), (\ref{eq:splitting}), and (\ref{eq:splitsquares}) we get
\begin{align}
&\frac{1}{2}\,\big|\d\tilde{\ps}\lrcorner\ps\big|^2+\frac{1}{8}\,\big|\d\tilde{\p}\lrcorner\ps\big|^2-\frac{1}{2}\,\big|\d\tilde{\p}\big|^2-\frac{1}{2}\,\big|\d\tilde{\ps}\big|^2-12\,\big|\d A\big|^2~=~\frac{1}{32}\big|\d J^2\lrc(v\w J^2)\big|^2\\
&~-\frac{1}{2}\,\big|\Re\,\d\Omega\lrc v\big|^2-\frac{1}{8}\big|\d J^2\lrc v\big|^2-\frac{1}{2}\big|\d v_\perp\big|^2+2\,\d v\lrc(a_2\w v)-2\,\Im\Omega\lrc\big((\d v\lrc v)\w \d v_\perp\big)\nonumber\\
&~+\frac{1}{8}\Im[ \d\Omega\lrc(v\w\bar{\Omega})]\times\Big[\frac{1}{4}\Im[ \d\Omega\lrc(v\w\bar{\Omega})]-2\d\tilde{v}_{\perp}\lrc J\Big]-\,\Re\,\d\Omega\lrc(v\w a_2\w J)-12\,\,\big|a_2\big|^2\nonumber\\
&~+\big(\Re\,\d\Omega\lrc(v\w J)\big)\lrc\Big[\frac{1}{2}\Re\,\d\Omega\lrc(v\w J)-\Im\Omega\lrc\d v_{\perp}\Big]+7\,\Im\Omega\lrc(a_2\w \d v_\perp)-18\,\big|\d A\lrc v\big|^2~,\nonumber
\end{align}
which gives exactly (\ref{eq:CYLimit}) when inserted into (\ref{eq:Rintermediate}). This confirms that our formulas are correct.

The bulk potential in this Calabi-Yau limit reads then
\begin{align}
V_0\,=\,&\frac{1}{4\kappa^2}\int\limits_{M}\text{dvol}_7\,e^{4A}\Big\lbrace\big|\d\tilde{J}-2\ast G\big|^2-3\,\big|\d\tilde{J}\lrc(v\w J)-6\,\d A\lrc v\big|^2+\big|\d\tilde{v}\big|^2+40|a_2|^2\\
&+\big|\Re\,\d\tilde{\Omega}\lrc v\big|^2+\big|\d\tilde{v}_\perp\lrc J\big|^2-\big|8\,a_2-\d\tilde{v}\lrc v\big|^2+\big|4\,a_2-\d\tilde{v}\lrc v+\Im\Omega\lrc\d\tilde{v}_\perp\big|^2\nonumber\\
&-\big|\frac{1}{4}\Im\big[\d\tilde{\Omega}\lrc(v\w\bar{\Omega})\big]\big|^2-\big|\Re\,\d\tilde{\Omega}\lrc(v\w J)+4\,a_2-\d\tilde{v}\lrc v+\Im\Omega\lrc\d\tilde{v}_\perp\big|^2\Big\rbrace~.\nonumber
\end{align}
As we have explained at the beginning of this section this potential is of BPS-like form. This becomes clear from (\ref{eq:TorsionConditions}) which states that for $W_1=\ldots=W_5=0$ all torison classes except $\Re E$ and $T_2$ have to vanish under SUSY and that $\Re E=-3\d A\lrc v$. This means that $a_2=0$ and that $\d \tilde{J}\lrc(v\w J)=6\d A\lrc v$ under SUSY, respectively. So we see that all squares vanish for a supersymmetric setting. Furthermore, the only component of the four-form flux $G$ that is not zero is $F^{2,2}$. This is in accordance with the necessity of a non-vanishing $F$ in Horava-Witten theory. But it is also consistent with the fact that one expects zero $H$-flux once one reduces the theory to a heterotic compactification on a Calabi-Yau manifold. In order to see how this precisely works we will consider the ten-dimensional limit in the next section. 
\subsection{The ten-dimensional limit}
\label{sec:10dlimit}
The most important consistency check of our previous results is the reduction of M-Theory to the heterotic string sector. The reduction is obtained by first performing the standard reduction of M-theory to type IIA theory as described e.g. in \cite{Becker:2007zj,Polchinski:1998rr} and then taking the limit $\pi\rho\rightarrow 0$ to move the two hyperplanes that are supporting the gauge fields on top of each other. This procedure should eventually lead to the results found in \cite{Held:2010az}. The eleven dimensional metric is then given by
\begin{equation}
\label{eq:metricMten}
\d s_{11}^2\,=\,e^{-\frac{2}{3}\Phi}\d s_{10}^2\,+\,e^{\frac{4}{3}\Phi}(\d x^{11}\,+\,C_1)^2\,=\,\,e^{2A'-\frac{2}{3}\Phi}(\d s'_{4})^{2}\,+\,e^{-\frac{2}{3}\Phi}(\d s'_{6})^{2}\,+\,e^{\frac{4}{3}\Phi}(\d x^{11}\,+\,C_1)^2~.
\end{equation}
Here $\Phi$ is the ten dimensional dilaton and $A'$ is the warp factor belonging to a compactification of string theory to four dimensions. $g'_4$ denotes the metric of the emerging four dimensional space, $g'_6$ of the compact six dimensional one, respectively, and $C_1$ is a one-form potential. Since we want to compare M-Theory compactifications to warped Minkowski space we take $g'_4$ to be the Minkowski metric. Comparing (\ref{eq:metricMten}) with the previously defined metrics (\ref{eq:metricMto4}) and (\ref{eq:MetricNearBound}) we see that
\begin{align}
\label{eq:warpings}
&2A'~=~2A+\frac{2}{3}\,\Phi~=~2A+\sigma\,,&  &\d s_7^2~=~e^{-\frac{2}{3}\Phi}(\d s'_{6})^{2}\,+\,e^{\frac{4}{3}\Phi}(\d x^{11}\,+\,C_1)^2~,&
\end{align}
where we remind the reader that $\sigma$ was the field used to describe the metric at the boundary in section \ref{sec:TheScaPot}. This means that the seven dimensional space $M$ splits into a six dimensional base $B$ and a one dimensional fiber. Since locally every seven dimensional SU$(3)$ manifold can be decomposed into a six and a one dimensional part and since this one dimensional piece is distinguished by $v$, we can also write
\begin{equation}
\d s_7^2~=~\d s_6^2\,+\,v\otimes v~.
\end{equation}
Thus the metric $g_6$ that we used to construct the SU$(3)$ structure and the metric $g'_6$ appearing in (\ref{eq:metricMten}) are related by $g_6=e^{-2\Phi/3} g'_6$ which gives $J=e^{-2\Phi/3}J'$ and $\Omega=e^{-\Phi}\Omega'$. For the one form $v$ we get
\begin{equation}
v~=~e^{\frac{2}{3}\Phi}\,(\d x_{11}\,+\,C_1)~.
\end{equation}
From the SUSY conditions for $v$ (\ref{eq:DiffSUSYcond}a) it follows that
\begin{equation}
\d A\,+\,\frac{1}{3}\d\Phi~=~\d A'~=~0\,,~~~~~~~~~~~~~~~~~~\d C_1~=~0~,
\end{equation}
as it should be for the heterotic string. We also see that $\d v\lrc v=W_0=-2/3\,\d \Phi_\perp=-\d \sigma_\perp$, which justifies the choice we made for $\sigma$ in section \ref{sec:BouPot}.

Since $(\d A\lrc v)$ is not zero, these equations also imply that the dilaton does depend on the $v$-direction. This is not the case in the heterotic theory, and in order to see how this dependence vanishes, it is necessary to analyze the behavior of the flux $F$ when the two hyperplanes are moved together. From equation (\ref{eq:FourFlux}) we have
\begin{equation}
F~\propto~\,\Big\lbrace\epsilon(x^{11})\, K^{(1)}\,-\,\frac{x^{11}}{\pi\rho}(K^{(1)}\,+\,K^{(2)})\Big\rbrace~.
\end{equation}
For $\pi\rho\rightarrow 0$ one can approximate the average of $F$ over the eleventh dimension by
\begin{equation}
\lim\limits_{\pi\rho\rightarrow 0}\,\langle F \rangle~=~\lim\limits_{\pi\rho\rightarrow 0}\,\frac{\int\limits_0^{\pi\rho}\d x^{11}\,e^{2\Phi/3} F}{\int\limits_0^{\pi\rho}\d x^{11}\,e^{2\Phi/3}}~\cong~\lim\limits_{\pi\rho\rightarrow 0}\,\frac{e^{2\Phi(0)/3}\int\limits_0^{\pi\rho}\d x^{11}\, F}{e^{2\Phi(0)/3}\int\limits_0^{\pi\rho}\d x^{11}}~\propto~\lim\limits_{\pi\rho\rightarrow 0}\,(K^{(1)}\,-\,K^{(2)})~.
\end{equation}
When the two hyperplanes are put on top of each other $K^{(1)}=K^{(2)}$ and hence $F=0$. For this reason we conclude that we can also set $(\d A\lrc v)$ to zero, once we go to the heterotic limit. A similar reasoning shows that $\d \tilde{J}\lrc v=\d\tilde{\Omega}\lrc v=0$ for $\pi \rho\rightarrow 0$.

The supersymmetry conditions of \cite{Held:2010az} can then be re-derived from our results (\ref{eq:DiffSUSYcond})
\begin{align}
\label{eq:relationSUSYcond}
&e^{-4A}\d\big(e^{4A}\,J\big)~=~2\,\ast G~=~2\,\ast\big(v\w H\big)&&\Rightarrow&&e^{-4A'+2\Phi}\d\big(e^{4A'-2\Phi}\,J'\big)~=~2\,\ast'_6 H&\nonumber\\
&e^{-2A}\d\big(e^{2A}\,J\w J\big)~=\,-4\,v\w G~=~0&&\Rightarrow&&e^{-2A'+2\Phi}\d\big(e^{2A'-2\Phi}\,J'\w J'\big)~=~0&\\
&e^{-3A}\d\big(e^{3A}\Omega\big)~=~0&&\Rightarrow&&e^{-3A'+2\Phi}\d\big(e^{3A'-2\Phi}\Omega'\big)~=~0~.&\nonumber
\end{align}
Note that in the first line one also has to rewrite the Hodge star, $\ast(v\w H)=e^{-2/3\Phi}\ast'_6 H$, and that we have to rescale $H$ by a factor of $-2$ in order to find complete agreement with \cite{Held:2010az}, due to our conventions. This shows that our SUSY conditions are indeed compatible with the SUSY conditions found for string theory compactifications on six dimensional SU$(3)$ structure manifolds.

Due to the restrictions $dA\lrc v=\d\tilde{J}\lrc v=\d\tilde{\Omega}\lrc v=0$ in the heterotic limit the linear piece (\ref{eq:linearpieces}) is identically zero and the bulk potential simplifies to 
\begin{equation}
\begin{split}
V_0\,=&\,\frac{1}{4\,\kappa_{10}^2}\int\limits_{B}\text{dvol}'_6\,e^{4A'-2\Phi}\,\Big\lbrace|\d \tilde{J}'-2\,\ast'_6 H|^2+\,|\d\tilde{\Omega}'|^2-\frac{1}{4}\,\big|\d\tilde{\Omega}'\lrcorner J'^2\big|^2+\frac{1}{4}\,|\d \tilde{J}'^2|^2\\
&- \,\Big|\frac{1}{4}\,\d\tilde{J}'^2\lrcorner J'^2-\frac{1}{2}\,\text{Re}(\d\tilde{\Omega}'\lrcorner\overline{\Omega}')+4\d A'\Big|^2+2\,\big|\d A'\big|^2\Big\rbrace~.
\end{split}
\end{equation}
All squares appearing in this formula are taken with respect to the metric $g'_6$ in order to get the right factors of $e^{\Phi}$ and we have absorbed the length of the eleventh dimension into the ten dimensional coupling $\kappa^2=2\pi\rho\,\kappa_{10}^2$. This is indeed the action to lowest order in $\alpha'$ as found in \cite{Held:2010az}.

The result for the boundary potential is even more easily obtained. Setting $3\sigma=2\Phi$ and adding the contributions of the two boundaries gives
\begin{align}
&V_b~=~\frac{\alpha'}{8\kappa_{10}^2}\int\limits_{B}\text{dvol}'_{6}\,e^{4A'-2\Phi}\, \Big(\text{Tr}|\calf^{(6)}_{\text{E}_8\times\text{E}_8}\lrc J|^2\,+\,2\,\text{Tr}|(\calf^{(6)}_{\text{E}_8\times\text{E}_8})^{2,0}|^2\Big)\\
&~~~~~-\frac{\alpha'}{8\kappa_{10}^2}\int\limits_{B}\text{dvol}'_{6}\,e^{4A'-2\Phi}\,\Big(\text{Tr}|R^{(6)'}_{+}\lrc J|^2\,+\,2\,\text{Tr}|(R^{(6)'}_{+})^{2,0}|^2\Big)\nonumber\\
&~~~~~-\frac{\alpha'}{8\kappa_{10}^2}\int\limits_{B}\text{dvol}'_{6}\,e^{4A'-2\Phi}\,\Big\lbrace 8\,e^{-2A'}\,(\nabla_i\nabla_j e^{A'})(\nabla^i\nabla^j e^{A'})-16\,\big|\d A'\lrc H\big|^2-12|\d A'|^4\Big\rbrace~.\nonumber
\end{align}
This is the $\mathcal{O}(\alpha')$ result of  \cite{Held:2010az} once the rescaling of $H$ is performed. We thus conclude that also in this limit our formulas provide the right results since we obtain exactly the scalar potential of the heterotic string compactified on an SU$(3)$ manifold.
\section{Conclusion}
In this paper we considered compactifications of Horava-Witten theory on seven dimensional manifolds with SU$(3)$ structure. In particular we shed light on the question whether it is possible to rewrite the bosonic action in a BPS-like form, by which we mean a form that is quadratic in the supersymmetry conditions and hence obey the equations of motion automatically once SUSY is imposed. 

The main obstacle in this analysis has been to rewrite the scalar curvature $R$ in a form that makes manifest its dependence on SUSY. We solved this problem by first rewriting $R$ in terms of the G$_2$ structure forms $\p$ and $\ps$ and then in terms of  the SU$(3)$ structure forms $v$, $J$, and $\Omega$.

Taking into account the warp factor $A$ and the Bianchi-identity of heterotic M-theory led to terms that vanished at most linearly under SUSY. We analyzed these terms and found that despite our expectation it was not possible to cancel all linear terms. \textit{So our conclusion is that it is in general not possible to put the bosonic action of Horava-Witten theory into BPS-like form}.

We confirmed our findings by crosschecking them in various limits. We showed that in the three limits of G$_2$ holonomy, six dimensional SU$(3)$ holonomy, and the reduction to the heterotic string, respectively, our equations are self-consistent and yield the expected results.

Since one of the main motivations of this work was to provide the domain wall SUSY-breaking scenarios of \cite{Held:2010az} with a strong coupling limit, one obvious question is how to lift DWSB-scenarios to M-theory in this context. Since the analysis of DWSB starts with a BPS-like background,  one should expect that also the M-theory lift has a BPS-like form. The simplest realization of this, that also allows for a general SU$(3)$ compactification of the heterotic string, is then given by a seven dimensional manifold with SU$(3)$ structure which satisfies $\d J\lrc v \propto J$ and $\d\Omega\lrc v\propto\Omega$. This means in particular that the torsion classes $S$ and $T_2$ are identically zero and hence no linearly vanishing terms survive. Having such a manifold one can then ask what happens if $\d\tilde\Omega=0$ does not longer hold, i.e. study the strong coupling limit of DWSB. However, this analysis is beyond the scope of this paper and will hopefully be given in another publication \cite{Held:toCome}.
\subsubsection*{Acknowledgments}
I would like to thank J.~Gray, S.~Groot Nibbelink, T.~Grimm, D.~H\"artl, S.~Halter, D.~L\"ust, L.~Martucci, F.~Marchesano, O.~Schlotterer, and T.~Rahn for fruitful discussion.
\newpage

\appendix

\section{Conventions}
\label{app:conventions}
There are various sorts of indices appearing in the paper. We denote with $M,N,\ldots$ all eleven dimensions and with $I,J,\ldots$ the first ten dimensions. $m,n,\ldots$ will be used for the seven internal dimensions and $i,j,\ldots$ for the first six internal dimensions. The Greek letters $\mu,\nu\ldots$ will stand for the four dimensional external space-time.

Our conventions on gamma matrices are as follows. $\Gamma^N$ denotes gamma matrices of SO$(1,10)$ which are $32\times32$ matrices. We split the $\Gamma^N$ according to
\begin{equation}
\Gamma^{\mu}=\gamma^{\mu}\otimes\mathds{1}=e^{-A}\hat{\gamma}^{\mu}\otimes\mathds{1}\;,~~~~~~\Gamma^m=\gamma_{(4)}\otimes\gamma^m~.
\end{equation}
The $\gamma^{\mu}$ are taken to be real and symmetric $4\times 4$ matrices, while $\gamma^m$ are purely imaginary, antisymmetric, and $8\times 8$. The four dimensional chirality operator is defined as 
\begin{equation}
\gamma_{(4)}=i\gamma^0\,\gamma^1\,\gamma^2\,\gamma^3~.
\end{equation}
From $\Gamma^{10}\Gamma^{10}=\mathds{1}$ and $\Gamma^{10}=\Gamma^{0}\cdot\ldots\cdot\Gamma^{9}$ it follows that 
\begin{equation}
\gamma^{10}\gamma^{10}=-i\,\gamma^4\ldots\gamma^{10}=\mathds{1}~. 
\end{equation}
An explicit representation of these gamma matrices can be found as in \cite{Gliozzi:1976qd} and is given by
\begin{align}
&\gamma^{0}\,=\,\left(\begin{array}{cc} 0 & \sigma^1\\-\sigma^1&0\end{array}\right)\,,\gamma^{1}\,=\,\left(\begin{array}{cc} 0 & \sigma^1\\\sigma^1&0\end{array}\right)\,,\gamma^{2}\,=\,\left(\begin{array}{cc} \sigma^3&0 \\0&\sigma^3\end{array}\right)\,,\gamma^{3}\,=\,\left(\begin{array}{cc} -\sigma^1&0 \\0&\sigma^1\end{array}\right)\,,\nonumber&\\
&&\\
&\gamma^{4,5,6}\,=\,\left(\begin{array}{cc} \alpha^{1,2,3}&0 \\0& \alpha^{1,2,3}\end{array}\right)\,,\gamma^{7,8,9}\,=\,\left(\begin{array}{cc} 0 & \beta^{1,2,3}\\ \beta^{1,2,3}&0\end{array}\right)\,,\gamma^{10}\,=\,\left(\begin{array}{cc} 0 & -i\,\mathds{1}_4\\ i\,\mathds{1}_4&0\end{array}\right)\,.\nonumber
\end{align}
With $\sigma^{i}$ the Pauli-matrices, the $4\times 4$ matrices $\alpha^i$, $\beta^j$ are given by
\begin{align}
&\alpha^1\,=\,\left(\begin{array}{cc} 0 & i\,\sigma^1\\-i\,\sigma^1&0\end{array}\right)\,, &&  \alpha^2\,=\,\left(\begin{array}{cc} 0 & -i\,\sigma^3\\i\,\sigma^3&0\end{array}\right)\,, &&  \alpha^3\,=\,\left(\begin{array}{cc} -\sigma^2 & 0\\0 &-\sigma^2\end{array}\right)\,,&\nonumber\\
&&\\
&\beta^1\,=\,\left(\begin{array}{cc} 0 & -\sigma^2\\-\sigma^2&0\end{array}\right)\,, &&  \beta^2\,=\,\left(\begin{array}{cc} 0 & i\,\mathds{1}\\-i\,\mathds{1}&0\end{array}\right)\,, &&  \beta^3\,=\,\left(\begin{array}{cc} \sigma^2 & 0\\0 &-\sigma^2\end{array}\right)\,.&\nonumber
\end{align}

Defining $\gamma^{m_1\ldots m_n}\equiv\gamma^{[m_1}\ldots\gamma^{m_n]}$ the relation between the antisymmetrized product of $n$ and $7-n$ gamma matrices is
\begin{equation}
\gamma^{m_1\ldots m_n}~=~i\,(-1)^{1+n(n-1)/2}\,\frac{1}{(7-n)!}\;\e\indices{^{m_1\ldots m_n}_{m_{n+1}\ldots m_7}}\gamma^{m_{n+1}\ldots m_7}~.
\end{equation}
Due to our manifestly real gamma matrices the Majorana condition on an 11 dimensional spinor $\e$ simply reads
\begin{equation}
\e^*~=~\e~.
\end{equation}
A slash will denote normalized antisymmetrized contraction with gamma matrices
\begin{equation}
\Sh{A_p}~=~\frac{1}{p!}\,\gamma^{n_1\ldots n_p}\,(A_p)_{n_1\ldots n_p}~,
\end{equation}
while $\lrc$ is used to contract forms
\begin{equation}
 A_p\lrc B_q~=~\frac{1}{(p-q)!\,q!}\,A_{n_1\ldots n_q m_1\ldots m_{p-q}}\,B^{n_1\ldots n_q}\d x^{m_1\ldots m_{p-q}}~.
\end{equation}
The formulas most frequently used for our calculations are
\begin{align}
&J_{mp}\,J^{pn}\,=\,-\delta_m^n\,+\,v_m\,v^n\,,& &\Omega_{mpq}\,\bar{\Omega}^{npq}\,=\,8\,(\delta_m^n\,+\,i\,J\indices{_m^n}\,-\,v_m\,v^n)\,,&\nonumber\\
&J_{mn}\,v^n\,=\,0\,=\,\Omega_{mnp}\,v^p\,,& &J\indices{_m^q}\,\Omega_{qnp}\,=\,-i\,\Omega_{mnp}&\\
&2\,\ast J\,=\,v\w J\w J\,,& &\ast\Omega\,=\,-\,i\,v\w\Omega\,.&\nonumber
\end{align}

\section{Scalar curvature}
\label{app:Curvature}
In this section we would like to show how to obtain (\ref{eq:scalcurv}) from  \cite{Bryant:2005mz}. For more details on the notation we refer to this paper. We start with their equation (4.27) $R=6\,\p_{mnp}\,T^{mnp}$, which can be rewritten as\footnote{We changed the $\e$-notation of \cite{Bryant:2005mz} such that $\e_{mnp}=\p_{mnp}$ and $\e_{mnpq}=\ps_{mnpq}$.}
\begin{equation}
\label{curv1}
R=6\,\p_{mnp}\,T^{mnp}=A\,\delta\tau_1+B\,\tau_0^2+C\,|\tau_1|^2+D\,|\tau_2|^2+E\,|\tau_3|^2~.
\end{equation}
The only two quantities not given in \cite{Bryant:2005mz} are $T_{mnp}$ and $\delta\tau_1$. To obtain $T_{mnp}$ one calculates the covariant derivative of the torison $\tau$ (not to be confused with the torsion classes) using (4.9) and (4.19) of \cite{Bryant:2005mz}
\begin{align}
D\tau_n&~=~\d\tau_n+\theta\indices{_n^m}\w\tau_m-\p\indices{_n^{mp}}\tau_p\w\tau_m\\
&~=~(\d T\indices{_n^m})\w\o_m+T\indices{_n^m}\,\d\o_m+\theta\indices{_n^m}\w\tau_m-\p\indices{_n^{mp}}\tau_p\w\tau_m\nonumber\\
&~=~(\d T\indices{_n^m} -T\indices{_n^p}\,\theta\indices{_p^m}+T\indices{_p^m}\,\theta\indices{_n^p})\w\o_m-(2\p\indices{_r^{mq}}\,T\indices{_n^r}\,T\indices{_q^p}+\p\indices{_n^{rq}}\,T\indices{_q^p}\,T\indices{_r^m})\o_p\w\o_m
\nonumber\\
&~=~(S\indices{_n^{mp}}-2\p\indices{_r^{mq}}\,T\indices{_n^r}\,T\indices{_q^p}-\p\indices{_n^{rq}}\,T\indices{_q^p}\,T\indices{_r^m})\o_p\w\o_m\nonumber\\
&~=~\frac{1}{2}\,T\indices{_n^{mp}}\,\o_p\w\o_m\nonumber
\end{align}
and hence
\begin{equation}
T_{mnp}~=~-2\,S_{mnp}-4\,\p_{qpr}\,T\indices{_m^q}\,T\indices{^r_n}-2\,\p_{mqr}\,T\indices{^q_p}\,T\indices{^r_n}~.
\end{equation}
From this we obtain for $R$
\begin{align}
\label{curv2}
R&~=~-12\,\p^{mnp}(S_{mnp}+2\,\p_{qpr}\,T\indices{_m^q}\,T\indices{^r_n}+\,\p_{mqr}\,T\indices{^q_p}\,T\indices{^r_n})\\
&~=~-12\,\p^{mnp}\,S_{mnp}+36\,(T\indices{_n^n})^2+12\,\ps^{mnpq}\,T_{mn}\,T_{pq}-24\,T^{mn}\,T_{mn}-12\,T^{mn}\,T_{nm}~.\nonumber
\end{align}
For $\delta\tau_1$ we get
\begin{align}
\delta\tau_1~=~-\ast~\d\ast\tau_1~=~-\p^{mnp}\,S_{mnp}-2\,\ps^{mnpq}\,T_{mn}\,T_{pq}-2\,T^{mn}(T_{mn}-T_{nm})~.
\end{align}
The torsion classes, defined through
\begin{equation}
\d\p~=~\tau_0\,\ps+3\,\tau_1\w\p+\ast\,\tau_3\;,~\d\ps~=~4\,\tau_1\w\ps+\tau_2\w\p~,
\end{equation}
are given by
\begin{align}
&\tau_0~=~\frac{1}{7}\,\d\p\lrcorner\ps~=~\frac{24}{7}\,T\indices{_n^n}\\
&\tau_1~=~-\frac{1}{12}\,\d\p\lrcorner\p~=~\frac{1}{12}\,\d\ps\lrcorner\ps~=~\p^{mnp}\,T_{mn}\,\o_p\nonumber\\
&\tau_2~=~\frac{1}{2}\left(\d\ps\lrcorner\p-\ast\,\d\ps\right)-2\,\tau_1\lrcorner\p~=~-\ast\,\d\ps+4\,\tau_1\lrcorner\p~=~\left(-\ps^{mnpq}\,T_{mn}+4\,T^{pq}\right)\,\o_p\w\o_q\nonumber\\
&\tau_3~=~\ast\d\p-\tau_0\,\p+3\,\tau_1\lrcorner\ps~=~\left(\frac{3}{7}\,T\indices{_q^q}\,\p^{mnp}-\frac{3}{2}\,\p^{qnp}(T\indices{_q^m}+T\indices{^m_q})\right)\,\o_m\w\o_n\w\o_p~.\nonumber
\end{align}
The squares of these are
\begin{align}
&\tau_0^2~=~\frac{576}{49}\,(T\indices{_n^n})^2\\
&|\tau_1|^2~=~\ps^{mnpq}\,T_{mn}T_{pq}+T^{mn}\left(T_{mn}-T_{nm}\right)\nonumber\\
&|\tau_2|^2~=~-12\,\ps^{mnpq}T_{mn}T_{pq}+24\,T^{mn}\left(T_{mn}-T_{nm}\right)\nonumber\\
&|\tau_3|^2~=~-\frac{72}{7}\,(T\indices{_n^n})^2+36\,T^{mn}\left(T_{mn}+T_{nm}\right)~.\nonumber
\end{align}
Comparing terms containing $S_{mnp}$, $(T\indices{_n^n})^2$, $\ps^{mnpq}\,T_{mn}\,T_{pq}$, $T^{mn}\,T_{mn}$, and $T^{mn}\,T_{nm}$, respectively, in \eqref{curv1} and \eqref{curv2} one gets equation (4.28) of \cite{Bryant:2005mz}
\begin{equation}
R~=~12\,\delta\tau_1+\frac{21}{8}\,\tau_0^2+30\,|\tau_1|^2-\frac{1}{2}\,|\tau_2|^2-\frac{1}{2}\,|\tau_3|^2~.
\end{equation}
Using (4.16) of \cite{Bryant:2005mz}, $\d\p=\ps^{mnpq}\,\tau_m\w\o_n\w\o_p\w\o_q$ and $\d\ps=-6\tau^p\w\o_p\w\p$, it is possible to show that
\begin{align}
&|\d\p|^2=36\,\left[2\,(T\indices{_n^n})^2+2\,T^{mn}\,T_{mn}+\ps^{mnpq}\,T_{mn}\,T_{pq}\right]\\
&|\d\ps|^2=36\,\left[2T^{mn}\left(T_{mn}-T_{nm}\right)+\ps^{mnpq}\,T_{mn}\,T_{pq}\right]~.\nonumber
\end{align}
From these expressions we find
\begin{align}
&|\tau_2|^2=|\d\ps|^2-48\,|\tau_1|^2\\
&|\tau_3|^2=|\d\p|^2-36\,|\tau_1|^2-7\,|\tau_0|^2~,\nonumber
\end{align}
leading to the final expression for the scalar curvature
\begin{align}
R~&=~12\,\delta\tau_1+\frac{49}{8}\tau_0^2+72\,|\tau_1|^2-\frac{1}{2}\,|\d\p|^2-\frac{1}{2}\,|\d\ps|^2\\
&=~-\nabla^m\left(\d\ps\lrcorner\ps\right)_m+\frac{1}{2}\,|\d\ps\lrcorner\ps|^2+\frac{1}{8}\,|\d\p\lrcorner\ps|^2-\frac{1}{2}\,|\d\p|^2-\frac{1}{2}\,|\d\ps|^2~.\nonumber
\end{align}
\newpage
\section{SUSY constraints}
\label{app:SUSYconstraints}
In this section we give the full list of constraints coming from the external SUSY variation (\ref{eq:exSUSYcond}). In these tables 'Ext' stands for equation (\ref{eq:exSUSYcond}) and $\gamma_{[n]}$ denotes $n$ antisymmetrized gamma matrices.
\begin{table}[ht!]
\centering
\label{constrainttable1a}
\caption{Constraints from $\delta\Psi_{\mu}=0$ coming from $\eta^{T}\gamma_{[n]}$Ext+Ext$^{T}\gamma_{[n]}\eta$.}
\renewcommand{\arraystretch}{1.3}
\begin{tabular}{p{0.9\textwidth}}
\hline
$72 e^{-A} w_0 \Sigma_0+(\tilde{\Sigma}_4)_{l_1l_2l_3l_4} G^{l_1l_2l_3l_4}=0$\hfill\\
$4 i \mu (\tilde{\Sigma}_3)_{mnp}+6e^{-A} w_0 (\Sigma_3)_{mnp}=6 (\tilde{\Sigma}_4)_{l_1mnp} \partial^{l_1}A+3(\tilde{\Sigma}_3)_{a_1a_2[m} G\indices{^{a_1a_2}_{np]}}$\\
$2 i \mu (\tilde{\Sigma}_4)_{mnpq}+3 e^{-A} w_0 (\Sigma_4)_{mnpq}+12 (\tilde{\Sigma}_3)_{[mnp} \partial_{q]}A=3 (\tilde{\Sigma}_4)_{l_1l_2[mn} G\indices{^{l_1l_2}_{pq]}}$\\
$3e^{-A}w_0(\Sigma_7)_{mnpqrst}+35(\tilde{\Sigma}_3)_{[mnp} G_{qrst]}=0$\\
\hline
\end{tabular}
\renewcommand{\arraystretch}{1.0}
\end{table}
\begin{table}[ht!]
\centering
\label{constrainttable1b}
\caption{Constraints from $\delta\Psi_{\mu}=0$ coming from $\eta^{T}\gamma_{[n]}$Ext-Ext$^{T}\gamma_{[n]}\eta$.}
\renewcommand{\arraystretch}{1.3}
\begin{tabular}{p{0.9\textwidth}}
\hline
$18 e^{-A}w_0 (\Sigma_1)_m+(\tilde{\Sigma}_3)^{l_1l_2l_3} G_{l_1l_2l_3m}=0$\\
$9 (\tilde{\Sigma}_3)_{l_1mn} \partial^{l_1}A+(\tilde{\Sigma}_4)_{l_1l_2l_3[m} G\indices{^{l_1l_2l_3}_{n]}}=9 e^{-A} w_0 (\Sigma_2)_{mn}$\\
$15 (\tilde{\Sigma}_4)_{[mnpq} \partial_{r]}A+10 (\tilde{\Sigma}_3)_{l_1[mn} G\indices{^{l_1}_{pqr]}}=3 e^{-A} w_0 (\Sigma_5)_{mnpqr}$\\
$3 e^{-A} w_0 (\Sigma_6)_{mnpqrs}+20 (\tilde{\Sigma}_4)_{l_1[mnp} G\indices{^{l_1}_{qrs]}}=0$\\
\hline
\end{tabular}
\renewcommand{\arraystretch}{1.0}
\end{table}
\begin{table}[ht!]
\centering
\label{constrainttable2a}
\caption{Constraints from $\delta\Psi_{\mu}=0$ coming from $\eta^{\dagger}\gamma_{[n]}$Ext+Ext$^{\dagger}\gamma_{[n]}\eta$.}
\renewcommand{\arraystretch}{1.3}
\begin{tabular}{p{0.9\textwidth}}
\hline
$72 (\Sigma_1)_{l_1} \partial^{l_1}A+(\Sigma_4)_{l_1l_2l_3l_4} G^{l_1l_2l_3l_4}=0$\\
$72 \Sigma_0 \partial_{m}A+(\Sigma_5)_{ml_1l_2l_3l_4} G^{l_1l_2l_3l_4}=0$\\
$72 (\Sigma_3)_{mnl_1} \partial^{l_1}A+(\Sigma_6)_{mnl_1l_2l_3l_4} G^{l_1l_2l_3l_4}=12 G_{mnl_1l_2}(\Sigma_2)^{l_1l_2}$\\
$e^{-A} w^*_0 (\tilde{\Sigma}^*_3)_{mnp}-e^{-A} w_0(\tilde{\Sigma}_3)_{mnp}+6 (\Sigma_2)_{[mn} \partial_{p]}A+\frac{1}{36} (\Sigma_7)_{mnpl_1l_2l_3l_4} G^{l_1l_2l_3l_4}$\\
$~~~=(\Sigma_3)\indices{^{l_1l_2}_{[m}} G_{np]l_1l_2}$\\
$e^{-A} w_0(\tilde{\Sigma}^*_4)_{mnpq}+e^{-A} w_0^*(\tilde{\Sigma}_4)_{mnpq}+2 (\Sigma_5)_{l_1mnpq} \partial^{l_1}A+\frac{2}{3} G_{mnpq}\Sigma_0$\\
$~~~=2 (\Sigma_4)_{l_1l_2[mn} G\indices{^{l_1l_2}_{pq]}}$\\
$3 (\Sigma_4)_{[mnpq} \partial_{r]}A+(\Sigma_1)_{[m} G_{npqr]}=(\Sigma_5)_{l_1l_2[mnp} G\indices{^{l_1l_2}_{qr]}}$\\
$5 (\Sigma_6)_{l_1l_2[mnpq} G\indices{^{l_1l_2}_{rs]}}=2 (\Sigma_7)_{l_1mnpqrs} \partial^{l_1}A+10 (\Sigma_2)_{[mn} G_{pqrs]}$\\
$6 (\Sigma_6)_{[mnpqrs} \partial_{t]}A+10 (\Sigma_3)_{[mnp} G_{qrst]}=3 (\Sigma_7)_{l_1l_2[mnpqr} G\indices{^{l_1l_2}_{st]}}$\\
\hline
\end{tabular}
\renewcommand{\arraystretch}{1.0}
\end{table}
\begin{table}[ht!]
\centering
\label{constrainttable2b}
\caption{Constraints from $\delta\Psi_{\mu}=0$ coming from $\eta^{\dagger}\gamma_{[n]}$Ext-Ext$^{\dagger}\gamma_{[n]}\eta$.}
\renewcommand{\arraystretch}{1.3}
\begin{tabular}{p{0.9\textwidth}}
\hline
$\mu\Sigma_0=0$\\
$\frac{4}{3} i \mu (\Sigma_1)_{m}=2 (\Sigma_2)_{l_1m} \partial^{l_1}A+\frac{1}{9} (\Sigma_3)^{l_1l_2l_3} G_{l_1l_2l_3m}$\\
$6 i \mu (\Sigma_2)_{mn}+18 (\Sigma_1)_{[m} \partial^{n]}A+(\Sigma_4)_{l_1l_2l_3[m} G\indices{^{l_1l_2l_3}_{n]}}=0$\\
$4 i \mu (\Sigma_3)_{mnp}+2 (\Sigma_1)^{l_1} G_{l_1mnp}$\\
$~~~=3 e^{-A} w_0 (\tilde{\Sigma}^*_3)_{mnp}+3 e^{-A} w_0^*(\tilde{\Sigma}_3)_{mnp}+6 (\Sigma_4)_{l_1mnp} \partial^{l_1}A+(\Sigma_5)_{l_1l_2l_3[mn} G\indices{^{l_1l_2l_3}_{p]}}$\\
$9 e^{-A} w_0 (\tilde{\Sigma}^*_4)_{mnpq}+12 i \mu (\Sigma_4)_{mnpq}+72 (\Sigma_3)_{[mnp} \partial_{q]}A+4 (\Sigma_6)_{l_1l_2l_3[mnp} G\indices{^{l_1l_2l_3}_{q]}}$\\
$~~~=9 e^{-A} w_0^*(\tilde{\Sigma}_4)_{mnpq}+24 (\Sigma_2)_{l_1[m} G\indices{^{l_1}_{npq]}}$\\
$12 i \mu (\Sigma_5)_{mnpqr}+60 (\Sigma_3)_{l_1[mn} G\indices{^{l_1}_{pqr]}}$\\
$~~~=18 (\Sigma_6)_{l_1mnpqr} \partial^{l_1}A+5 (\Sigma_7)_{l_1l_2l_3[mnpq} G\indices{^{l_1l_2l_3}_{r]}}$\\
$i \mu (\Sigma_6)_{mnpqrs}+9 (\Sigma_5)_{[mnpqr} \partial_{s]}A=10 (\Sigma_4)_{l_1[mnp} G\indices{^{l_1}_{qrs]}}$\\
$2 i \mu (\Sigma_7)_{mnpqrst}+35 (\Sigma_5)_{l_1[mnpq} G\indices{^{l_1}_{rst]}}=0$\\
\hline
\end{tabular}
\renewcommand{\arraystretch}{1.0}
\end{table}
\newpage
\vfill
~~

\end{document}